\documentclass[twocolumn,showpacs,prb,10pt]{revtex4-1}
\usepackage[english]{babel}
\usepackage{amsmath,graphicx,amssymb} 
\usepackage{times}
\usepackage{epstopdf} 
\usepackage{epsfig}
\usepackage{placeins} 

\newcommand{\eq}[1]{\begin{equation}#1\end{equation}}
\newcommand{\eqa}[1]{\begin{eqnarray}#1\end{eqnarray}}
\newcommand{\RE}{{\rm Re}}
\newcommand{\IM}{{\rm Im}}
\newcommand{\im}{{\rm i}}

\begin{document} 
\title{Transport properties of the metallic state of overdoped cuprate superconductors
 from an anisotropic marginal Fermi liquid model}

\author{J. Kokalj$^1$}
\email{j.kokalj@uq.edu.au}
\thanks{On leave from J.\ Stefan Institute, Ljubljana, Slovenia}
\author{N. E. Hussey$^2$}
\author{Ross H. McKenzie$^1$}
\affiliation{$^1$School of Mathematics and Physics, University of Queensland,
  Brisbane, 4072 Queensland, Australia} 
\affiliation{$^2$H. H. Wills Physics Laboratory, University of Bristol,
  Tyndall Avenue, BS8 1TL, United Kingdom} 

\date{\today}

\begin{abstract}
  We consider the implications of a phenomenological model self-energy
  for the charge transport properties of the metallic phase of the
  overdoped cuprate superconductors.  The self-energy is the sum of
  two terms with characteristic dependencies on temperature,
  frequency, location on the Fermi surface, and doping.  The first
  term is isotropic over the Fermi surface, independent of doping, and
  has the frequency and temperature dependence characteristic of a
  Fermi liquid.  The second term is anisotropic over the Fermi surface
  (vanishing at the same points as the superconducting energy gap),
  strongly varies with doping (scaling roughly with $T_c$, the
  superconducting transition temperature), and has the frequency and
  temperature dependence characteristic of a marginal Fermi liquid.
  Previously it has been shown this self-energy can describe a range
  of experimental data including angle-dependent magnetoresistance
  (ADMR) and quasi-particle renormalisations determined from specific
  heat, quantum oscillations, and angle-resolved photo-emission
  spectroscopy (ARPES).  Without introducing new parameters and
  neglecting vertex corrections we show that this model self-energy
  can give a quantitative description of the temperature and doping
  dependence of a range of reported transport properties of Tl2201
  samples.  These include the intra-layer resistivity, the frequency
  dependent optical conductivity, the intra-layer magnetoresistance,
  and the Hall coefficient.  The 
  temperature dependence of the latter two are particularly sensitive to
  the anisotropy of the scattering rate and to the shape of the Fermi
  surface.  In contrast, the temperature dependence of the Hall angle
  is dominated by the Fermi liquid contribution to the self-energy
  that determines the scattering rate in the nodal regions of the
  Fermi surface.

\end{abstract}

\maketitle 

\section{Introduction}

Much research on strongly correlated electron materials, with
high-temperature superconducting cuprates being the prominent example,
is focused on the experimental or theoretical determination of the
relevant electronic self-energy. That is because the self-energy can
provide insight into the underlying quantum many-body physics.  Proper
knowledge of the self-energy in the metallic phase of high-temperature
superconductors is also believed to be a step towards solving the
mystery of high-temperature superconductivity since ultimately
superconductivity is an instability in the metallic state.  A model
self-energy capable of a unified description of results from many
experiments is therefore desirable and provides a benchmark for
comparison with microscopic theories based on lattice effective
Hamiltonians such as the $t-J$ and Hubbard models.

In the last two decades experimental data has been used to deduce the
self-energy, both directly and indirectly.  Angle-resolved
photoemission spectroscopy (ARPES) offers information on both the real
and imaginary part of the self-energy \cite{damascelli03, valla99,
  valla00, kordyuk04, kaminski05, chang08, zhu08, plate05}, specific
heat\cite{wade94,loram94} provides information through renormalization
effects, and angle-dependent magnetoresistance
\cite{abdel06,abdel07,analytis07,kennett07} (ADMR) provides
information on the imaginary part of the self-energy or scattering
rate close to the Fermi level.  Further information about the
temperature and doping dependence can be obtained from measurements of
the resistivity \cite{tyler,tyler98,manako92,mackenzie96,hussey08},
intra-layer magnetoresistance \cite{hussey96}  and
Hall effect \cite{manako92,mackenzie96,hussey08,kubo91,hussey96}.  In
addition, the optical conductivity
\cite{basov05,puchkov96,ma06,basov11} provides information on the
frequency dependence of the self-energy.

Previous work \cite{kokalj11} introduced a particular model
self-energy, motivated by ADMR \cite{abdel06,abdel07}, that could
describe consistently and quantitatively ADMR and a number of
quantities determined by the real part of the self-energy, including
ARPES dispersion \cite{plate05}, specific heat \cite{wade94,loram94},
and effective masses deduced from quantum oscillations
\cite{rourke10}, in the entire overdoped regime of
Tl$_2$Ba$_2$CuO$_{6+\delta}$ (Tl2201).  In this paper we extend our
analysis to description of transport properties, which are largely
determined by the imaginary part of the self-energy. Properties
considered include the intra-layer resistivity
\cite{tyler,tyler98,manako92,mackenzie96}, Hall effect
\cite{manako92,mackenzie96,kubo91,hussey96}, intra-layer
magnetoresistance \cite{hussey96}, and optical conductivity
\cite{basov05,puchkov96,ma06}.  We show that all of these can be
quantitatively described with our model self-energy without any
additional fitting parameters for the entire overdoped regime for
Tl2201.

The outline of the paper is as follows.  In Section
\ref{sec_modelselfenergy} we review the
form of the self-energy and its parametrisation.  Section \ref{sec_IntralayerConductivity}
considers the DC conductivity and the frequency dependent
conductivity. It is shown that at high temperatures and frequencies
these are sensitive to the cut-off frequency which appears in the
Fermi liquid term in the self-energy.  In Section \ref{sec_Halleffect} we show that the
Hall coefficient strongly depends on the shape of the Fermi surface
and that its non-monotonic temperature dependence gives strong support
for our model self-energy. We also argue that the observed
non-monotonic temperature dependence of the Hall coefficient cannot be
captured with some alternative models, e.g., with the isotropic
marginal Fermi Liquid model \cite{abrahams03}. We also argue that the
observed $T^2$ dependence of the Hall angle arises because it is
dominated by the isotropic part of the self-energy, which also equals
the smallest scattering rate on the Fermi surface in the nodal
direction, and that the contribution of the anisotropic part is
suppressed. Hence, results on the Hall angle give additional support
to our model self-energy, in particular the $T^2$ dependence of the
isotropic part of self-energy. 
In Section \ref{sec_Intralayermagnetoresistance} we consider the
intra-layer magnetoresistance. 
 In Section \ref{sec_Comparisonwithmicroscopicmodels}
 we briefly review
relevant results from microscopic model calculations. Although,
several are qualitatively consistent with the model self-energy, they
tend to obtain scattering rates that are significantly less than
observed.  Section \ref{sec_Conclusions}  contains some conclusions
and suggestions for 
possible future work. 

\section{Model self-energy}
\label{sec_modelselfenergy}
Our model self-energy is motivated by the angle-dependent
magnetoresistance (ADMR) experiments on overdoped Tl2201
\cite{abdel06,abdel07,analytis07}, where two distinct scattering rates
were uncovered.  The first is more Fermi liquid (FL) like and is
isotropic over the Fermi surface, weakly doping dependent, and shows
$T^2$ dependence at low $T$. The second has a marginal Fermi
liquid\cite{varma89,littlewood91,varma02} frequency and temperature
dependence and is strongly anisotropic over the Fermi surface (the
same anisotropy as the superconducting gap). Its strength follows the
doping dependence of $T_c$ in the strongly overdoped regime and is
linear in $T$ down to the lowest $T$.

\begin{figure}[htb] 
\centering 
\includegraphics[width = 0.3\textwidth, angle=-90]
  {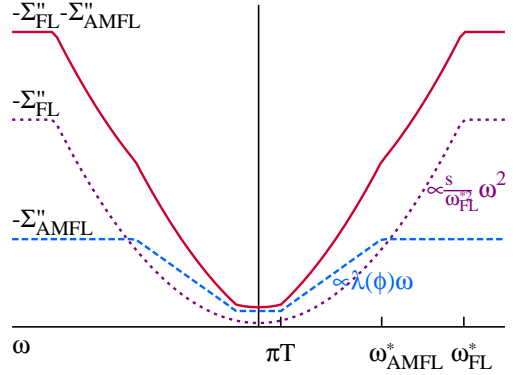}\\
  \caption{Frequency dependence
of the imaginary part of the model self-energy.
     The first part $\Sigma_\textrm{FL}$, is  Fermi liquid like,  with a
    quadratic $\omega$ and $T$ dependence, 
up to the high-$\omega$ cutoff $\omega_\textrm{FL}^{*}$,
and with a prefactor $s/\omega_\textrm{FL}^{*2}$,
 This part is taken to be doping
    independent. The second part is the anisotropic marginal
    Fermi liquid part, whose imaginary part is constant for
    $\omega<\pi T$ and proportional to $T$, while it is linear in
    $\omega$ for higher $\omega$ up to a high-$\omega$ cutoff
    $\omega_\textrm{AMFL}^*$. Its strength is given by   $\lambda(\phi)$,
    which is strongly anisotropic over the Fermi surface and also
    strongly doping dependent.}
\label{fig0}
\end{figure}

Accordingly, our model self-energy can be written,       
\eq{
\Sigma''({\bf k},\omega)=\Sigma''_\textrm{FL}(\omega)+\Sigma_\textrm{AMFL}''(\phi,\omega),
\label{eq_selfenergy}
} 
where $\phi$ denotes the position on the Fermi surface (see Fig. \ref{fig4}). The
imaginary part of the isotropic FL like self-energy 
is given by \cite{JackoNP09}
\begin{equation}
\Sigma_\textrm{FL}''(\omega)= \left \{ 
\begin{array}{ll}
  -\frac{1}{2\tau_0}-s\frac{\omega^2+\pi^2 T^2}{\omega_\textrm{FL}^{*2}}&\textrm{ for
  }\frac{\omega^2+\pi^2 T^2}{\omega_\textrm{FL}^{*2}}\leq 1 ,\\ 
  \big[-\frac{1}{2\tau_0}-s\big]F\big(\frac{\omega^2+\pi^2
    T^2}{\omega_\textrm{FL}^{*2}}\big)&\textrm{ for
  }
\frac{\omega^2+\pi^2 T^2}{\omega_\textrm{FL}^{*2}} > 1.
\end{array}
\right .
\label{sigmaflppomega}
\end{equation}
Here $1/(2\tau_0)$ accounts for the impurity scattering,
and Matthiessen's rule is implicitly assumed. The 
parameter $s$ gives the
strength of the FL like self-energy part and $\omega_\textrm{FL}^*$ is
the high-$\omega$ cutoff (see Fig. \ref{fig0}).  We use units
$\hbar=k_B=1$. $\Sigma_\textrm{FL}''$ is quadratic in $\omega$ and $T$
at low $\omega$ and $T$. The function $F$ is a slowly decreasing
function with $F(1)=1$, which we simply approximate with a constant. 

The anisotropic marginal Fermi liquid (AMFL) part of the self-energy (its
imaginary part) has the following form,
\begin{equation}
\Sigma_\textrm{AMFL}''(\phi,\omega)= \left \{ 
\begin{array}{ll}
\lambda(\phi)(- \frac{\pi}{2}x)&\textrm{ if } |\omega|\leq \omega_\textrm{AMFL}^* ,\\
\lambda(\phi)(- \frac{\pi}{2}\omega_\textrm{AMFL}^*)&\textrm{ if }
|\omega| > \omega_\textrm{AMFL}^* , 
\end{array}
\right .
\label{simgamflppphiomega}
\end{equation}
where $\lambda(\phi)$ determines its strength and is anisotropic over
the Fermi surface, $x=\max(|\omega|,\pi T)$, and $\omega_\textrm{AMFL}^*$ is the
high-$\omega$ cutoff. $\Sigma_\textrm{AMFL}''(\phi,\omega)$ is linear
in $\omega$ and $T$ for low $\omega$ or low $T$ (see Fig. \ref{fig0}).
The real part of the self-energy is obtained 
from a Kramers-Kronig relation and is not
explicitly given here. Explicit low $\omega$ behaviour of the real
part can be found in Ref. \onlinecite{kokalj11}.

Parameters of the model self-energy were already extracted in
Ref. \onlinecite{kokalj11} for overdoped Tl2201. From ADMR one can
estimate that $s/\omega_\textrm{FL}^{*2} = 9.2 $ eV$^{-1}$ and
\begin{equation} 
\lambda(\phi) = 1.6
\cos ^2 (2 \phi) T_c(p)/T_c^{\max}, 
\end{equation}
where $T_c(p)$ is the doping ($p$) dependent transition temperature
and $T_c^{\max}$ is the maximal transition temperature.  For the
doping dependence of $T_c$ we use the phenomenological relation
\cite{tallon95} $T_c(p)/T_c^\textrm{max}=1-82.6(p-0.16)^2$ with
$T_c^\textrm{max}=93 $ K at the optimal doping $p=0.16$ for
Tl2201. This $T_c(p)$ relation is for illustrative purposes only since
the superconductivity actually survives up to $p=0.31$, as was found
in Ref. \onlinecite{bangura10}.  In addition, $\omega_\textrm{FL}^*$
was estimated\cite{kokalj11} to be 0.23 eV from specific heat
measurements in the highly overdoped and non-superconducting regime.
The cutoff $\omega_\textrm{AMFL}^*$ only weakly influences the results
\cite{kokalj11} because it only enters the real part of the self
energy via a logarithmic dependence and is here taken to be $0.2$
eV. For the Tl2201 samples used in ADMR \cite{abdel06,french09} the
impurity scattering rate was estimated to be, $1/(2\tau_0) \sim 4$
meV.  These parameter values for the model self-energy give a
consistent description of several experiments, including ADMR,
specific heat, quantum oscillations, and the quasi-particle dispersion
seen in ARPES \cite{kokalj11}.

For Hubbard models, there is an additional constraint on the self-energy,
 and in particular its
high-frequency behaviour, via the sum rule \cite{vilk97} 
\begin{equation}
\int d\omega(-\Sigma_\sigma''(k,\omega))=\pi U
n_{-\sigma}(1-n_{-\sigma}), 
\label{eq_sigmasumrule}
\end{equation}
with $U$ being the on site Coulomb interaction strength, and
$n_\sigma=(1+p)/2$ being the density of electrons with spin $\sigma$.
Our model self-energy does not obey this sum rule since
$\Sigma''(k,\omega)$ stays finite for $\omega\to \infty$. To fulfill
the sum rule our $\Sigma''(k,\omega)$ should be strongly suppressed at
high frequencies. We estimate this suppression should occur at $\omega
\sim 5$ eV (for $U=8t$). Such a suppression would not influence our
results, since they are determined by the value of the self energy at
much lower frequencies. Hence, we do not employ the suppression and
the self-energy sum rule in this work.  In contrast, in the next
Section it is shown that the behaviour of the self-energy near our
cutoff frequencies $\omega_\textrm{FL}^*$ and $\omega_\textrm{AMFL}^*$
does affect some observed transport properties at high temperatures
and frequencies.

For the bare band dispersion $\epsilon_0(k)$ we approximate the LDA
results in Ref. \onlinecite{peets07} with the following hopping
parameters. $t_1 = 0.438$, $t_2 = -0.150$, $t_3 = 0.084$, $t_4 =
-0.013$, $t_5 = -0.020$, $t_6 = 0.029$, all expressed in eV (for
details see Supplemental material of Ref. \onlinecite{kokalj11}).  To
obtain the Fermi surface volume of the overdoped regime we apply a
rigid band shift through the chemical potential $\mu$.  The main
doping dependence of our results does not come from the band filling,
which we therefore keep fixed, but rather from the doping (or $T_c$)
dependence of the self-energy.  Shifting the chemical potential from
values appropriate for highly overdoped to optimal doping (e.g., from
$p=0.3$ to 0.15) induces only a small change of our results (see
Fig. \ref{fig7} for example).

\section{Intra-layer Conductivity}
\label{sec_IntralayerConductivity}
The frequency dependent 
conductivity is approximated with the bubble diagram in which the
non-interacting Green's functions are exchanged with interacting ones
and vertex corrections are neglected \cite{bruus}.
\eqa{
&\RE & \ \sigma_{xx}(\omega) =
\frac{2\pi e^2}{V}\sum_{\vec{k}} v_{0,x}^2 (\vec{k}) \nonumber\\
&&\times\int dy \frac{n_F(y)-n_F(y+\omega)}{\omega}
 A(\vec{k},y) A(\vec{k},y+\omega), 
}
where $v_{0,x}(\vec{k})$ is the bare band velocity in the
$x$ direction at wave vector $\vec{k}$,
$n_F(y)$ is the Fermi-Dirac
 distribution function and $A(\vec{k},\omega)$ is the spectral
function
\eq{
A(\vec{k},\omega)=-\frac{1}{\pi} \IM G^R(\vec{k},\omega)
}
where $G^R(\vec{k},\omega)$ is the retarded Green's function. 

Our interest is mostly in low $T$ and low $\omega$ properties of the
conductivity for which the parameter space close to the Fermi surface
is the most relevant (mainly due to the factor
$(n_F(y)-n_F(y+\omega))/\omega$).  In this parameter space we can
linearize the bare-band dispersion 
\eq{
\epsilon_{0}(\vec{k})=\epsilon_0(k_r, \phi) \simeq \epsilon_F
+v_{0,F,r} (\phi)(k_F(\phi)-k_r), 
\label{eq_epsilon0k}
}
where $k_F(\phi)$ is a Fermi momentum at angle $\phi$, which is the 
 angle between the $(\pi,\pi)$-$(0,\pi)$ and $(\pi,\pi)$-$k$
 directions (Fig. \ref{fig4}). 
 $v_{0,F,r}(\phi)$ is the derivative of the bare
band dispersion in the $k_r$ direction [i.e., the radial from $(\pi,\pi)$, see
Fig. \ref{fig4}]. For a circular Fermi surface 
 $v_{0,F,r}(\phi)$ just corresponds to a  Fermi velocity.  

By performing the integral over $k_r$ the optical conductivity can
be approximated for a quasi 2D system with 
\eqa{
\RE \ \sigma_{xx}(\omega) = \frac{ e^2}{4 \pi^2 d}  
 \int d\phi
\int dy  \frac{n_F(y)-n_F(y+\omega)}{\omega}\nonumber\\
\times\frac{k_F(\phi) v_{0,F}^2(\phi)}{v_{0,F,r}(\phi)}
 \IM \frac{1}{\omega+\Sigma^R(\phi,y)-\Sigma^A(\phi,y+\omega)}.
\label{eq_resigxxomega}
}
$d$ is the distance between CuO layers ($d=11.6$ \AA\ for Tl2201
\cite{french09}), $v_{0,F}(\phi)$ is a Fermi velocity while
$\Sigma^R(\phi,\omega)$ and $\Sigma^A(\phi,\omega)$ are the retarded
and advanced self-energies, respectively.  We assume that they are
only $\phi$-dependent in $\vec{k}-$space (anisotropic over the Fermi surface) in
addition to our proposed $\omega$ and $T$ dependencies.  In deriving
Eq. (\ref{eq_resigxxomega}) the integral over $k_r$ was extended to
[$-\infty$,$\infty$] which is a good approximation for low $T$ and
$\omega$ due to the strongly peaked spectral function close to the
FS. This means any effects of van Hove singularities or band edges are
neglected.  Eq. (\ref{eq_resigxxomega}) can be viewed as a
generalization of Eq. (12) in Ref. \onlinecite{basov05} to the
case of a $\phi$ dependent self-energy.

The imaginary part of the optical conductivity can be obtained by
the Kramers-Kronig transformation,
\eq{
\IM \ \sigma_{xx}(\omega)=-\frac{1}{\pi} \mathcal{P} \int_{-\infty}^{\infty}
\frac{\RE \ \sigma_{xx}(\omega')}{\omega'-\omega}d\omega',
}
or by generalizing Eq. (\ref{eq_resigxxomega}) to the complex conductivity
\eqa{
\sigma_{xx}(\omega) = \frac{ \im e^2}{4 \pi^2 d}  
 \int d\phi
\int dy  \frac{n_F(y)-n_F(y+\omega)}{\omega}\nonumber\\
\times\frac{k_F(\phi) v_{0,F}^2(\phi)}{v_{0,F,r}(\phi)}
 \frac{1}{\omega+\Sigma^A(\phi,y)-\Sigma^R(\phi,y+\omega)}.
\label{eq_sigmaxx0}
}
The plasma frequency $\omega_p$ is determined by the high frequency behaviour
($\omega \gg$ band width), 
\eq{
\omega_p^2 \equiv \frac{1}{\epsilon_0} {\rm lim}_{\omega\to \infty} \ 
\omega
\IM  \ \sigma_{xx}(\omega)
\label{eq_plasma}
}
and in our case this quantity is given by   the
following integral over the Fermi surface
\eq{
\omega_p^2 = \frac{e^2}{4\pi^2 d \epsilon_0 } \int d\phi
\frac{k_F(\phi) v_{0,F}(\phi)^2}{v_{0,F,r}(\phi)}.
\label{eq_omegap2}
}
 The above expression is equivalent
to the band theory expression
\cite{ashcroft}, 
\eq{
  \omega_p^2=\frac{e^2}{\epsilon_0} \int \frac{d^3k}{4 \pi^3}
  n_F(\epsilon_0(k)) \frac{\partial^2\epsilon_0(k)}{\partial k_x^2}. 
}
This equivalence can be shown 
by integrating by  parts, confining the integral to the Fermi surface due to
derivative of a Fermi function and then using the symmetry
$\sigma_{xx}=\sigma_{yy}$.  Here $\epsilon_0$ is the static dielectric
constant.

Using these expressions with our bare band dispersion (see Section
\ref{sec_modelselfenergy}) we obtain $\omega_p \simeq 23000$ 
cm$^{-1}$, while in Ref. \onlinecite{puchkov96} they experimentally obtain
$\omega_p\sim 15 100$ cm$^{-1}$ by integrating $\RE \sigma(\omega)$ up
to $\sim 8000$ cm$^{-1}$.  We believe that this is not 
a high enough frequency to fully exhaust the sum rule.

\subsection{DC conductivity}

\label{sec_dcconductivity}
In the limit of $\omega\to 0$ further simplifications can be made,
\eq{
\frac{1}{\omega+\Sigma^R(\phi,y)-\Sigma^A(\phi,y+\omega)}\to
\frac{1}{2 \im \Sigma''(\phi,y)}, 
}
where $\Sigma''(\phi,y)$ stands for imaginary part of the retarded
self-energy. 
Furthermore,
\eq{
\frac{n_F(y)-n_F(y+\omega)}{\omega} \to 
-\frac{\partial n_F(y)}{\partial y}.
}
The DC conductivity can then be written as
\eqa{
\RE\sigma_{xx} &= &\frac{ e^2}{4 \pi^2 d}  
 \int d\phi
\frac{k_F(\phi) v_{0,F}^2(\phi)}{v_{0,F,r}(\phi)}\nonumber\\
&&\times \int dy  (-\frac{\partial n_F(y)}{\partial y})
 \frac{1}{-2\Sigma''(\phi,y)}.
\label{eq_dc_cond}
}

For the bare band dispersion appropriate to Tl2201 the
 pre-factor $\frac{k_f(\phi) v_{0,F}(\phi)^2}{v_{0,F,r}(\phi)}$
 turns out to be relatively
constant with $\phi$ (variation $<$20\%). In comparison
the  anisotropy
of the self-energy ($1/\Sigma''$ can vary by a factor of
more than two) and so the pre-factor can
therefore be taken out of the integral, 
replaced with its average value and expressed 
in terms of $\omega_p$ [Eq. (\ref{eq_omegap2})]. 
With this approximation we can rewrite the
expression for the frequency dependent conductivity in
Eq. (\ref{eq_sigmaxx0}) in a similar form to Equation (12)
in Ref. \onlinecite{basov05},
\eqa{
\sigma_{xx}(\omega) = \frac{\im \omega_p^2 \epsilon_0}{2\pi} 
\int dy  \frac{n_F(y)-n_F(y+\omega)}{\omega}\nonumber\\
\times
\int d\phi \frac{1}{\omega+\Sigma^A(\phi,y)-\Sigma^R(\phi,y+\omega)}.
\label{eq_sigmaxx}
}
Using the $\cos^2(2\phi)$ dependence of 
our model self-energy, one can perform the integral over $\phi$ in the
Eq. (\ref{eq_sigmaxx}) for $\omega=0$. At this 
point only the integral over the frequency $y$ remains, which can be to
the lowest order at low $T$ calculated with the use of 
\eq{
-\frac{\partial n_F(y)}{\partial y} \to
\delta(y).
\label{eq_partialnf}
}
This is a good approximation, if the self-energy (or $1/\Sigma''$) is
a fairly constant function of $\omega$ for $|\omega|\lesssim T$. 
However, further improvements can be made by expanding the self-energy
part to $y^2$ term and then numerically approximating the $y$ integral
by Pade approximation, which gives errors less than $10^{-6}$.

The resulting expression allows us to perform fits of the measured
resistivity ($\rho_{xx}=1/\RE \sigma_{xx}$) using the three main
parameters of our model: the strength of 
impurity scattering $1/(2\tau_0)$, the strength of AMFL part of self-energy
$\lambda$ [where $\lambda(\phi)\equiv \lambda \cos^2(2\phi)$],
 and the  strength of FL part of self-energy
$s/\omega_\textrm{FL}^{*2}$. 

\begin{figure}[htb] 
\centering 
 \includegraphics[width = 0.32\textwidth, angle=-90]
  {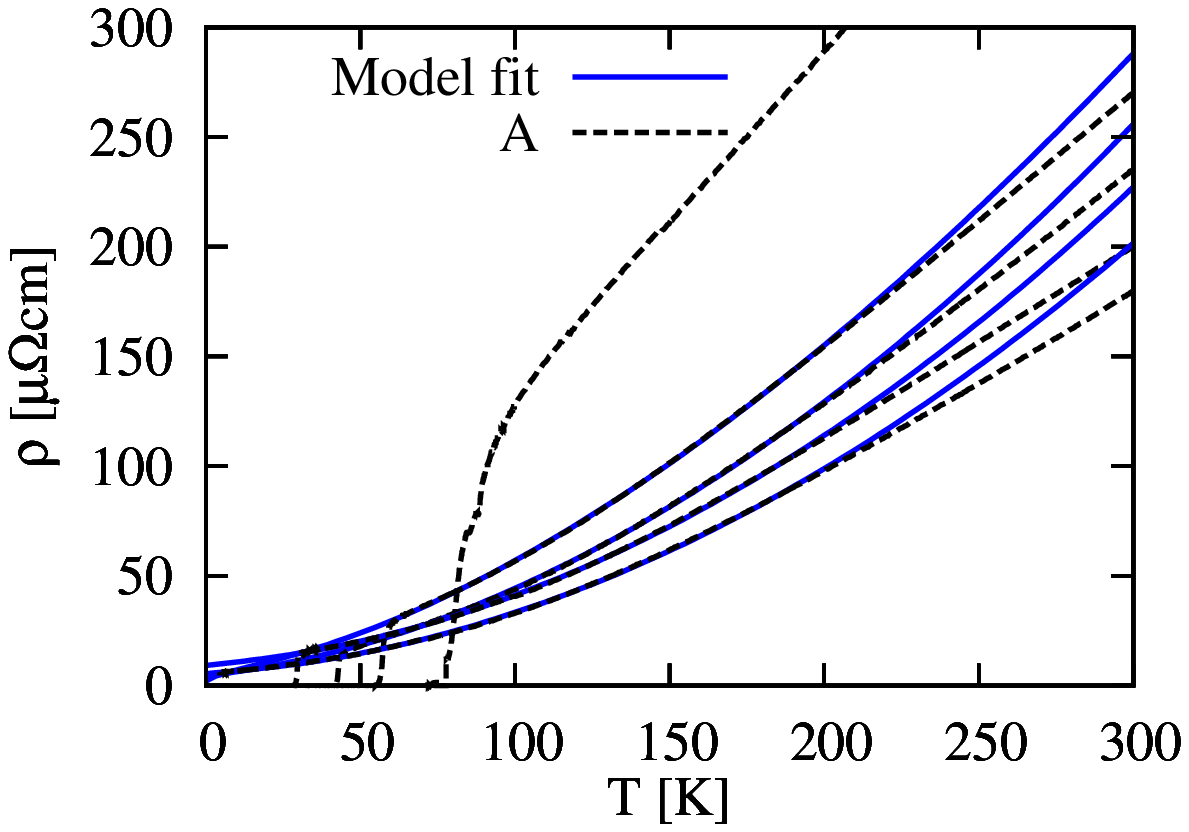}
 \includegraphics[width = 0.32\textwidth, angle=-90]
  {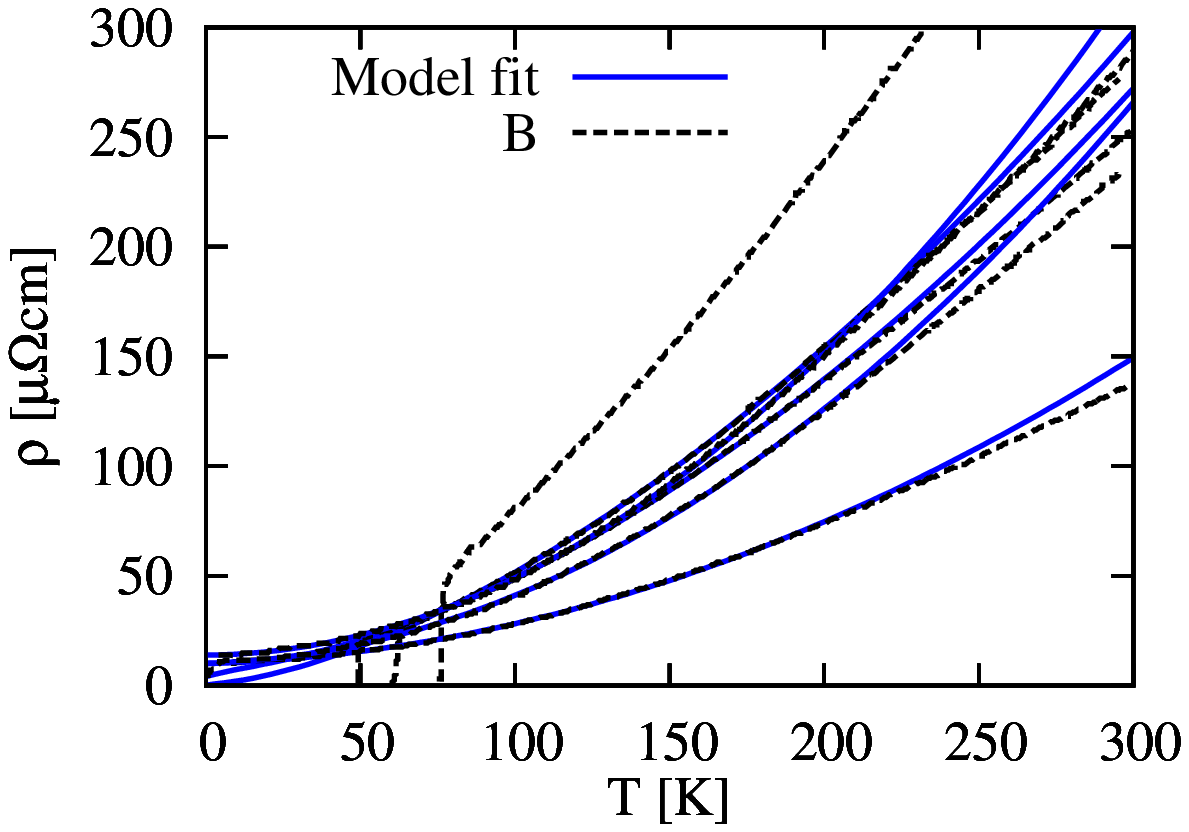}\\
\caption{
Comparison of the measured
 temperature dependence
of the intra-layer resistivity $\rho_{xx}$
to fits for the self-energy model.   
 Fits are performed in the $T$
  range from above $T_c$ to
  200 K and agree nicely with the data. At higher $T>200$ K our
  model predicts a stronger increase in $\rho_{xx}$, which could be improved
  by smoother high frequency cutoffs in the self-energy. This
  would introduce new parameters and is beyond the scope of this
  paper.  $T_c$ values and references for the data denoted with A and
  B are given in Table \ref{tab1}. 
}
\label{fig1}
\end{figure}

The resulting fits for various Tl2201 samples with different $T_c$s are
shown in Fig. \ref{fig1}.
\begin{table}[htb] 
\caption{Experimental data 
sets for the temperature
dependence of the DC intra-layer resistivity
which are fit to our self-energy model. (Compare Figures
\ref{fig1} and  \ref{fig2}).
The 
corresponding $T_c$'s and references are listed.}
\label{tab1}
\begin{center}
\begin{tabular}{ c c c }
\hline\noalign{\smallskip}
 \hline\noalign{\smallskip}
 Data identifier & Data $T_c$ [K]& Reference\\
\noalign{\smallskip}\hline\noalign{\smallskip}
A &  0, 30, 43, 57, 83 & A.W. Tyler et al., Ref. \onlinecite{tyler}\\
B & 0, 7, 10, 48, 63, 76 & T. Manako et al., Ref. \onlinecite{manako92}\\
C &  15 & A.P. Mackenzie et al., Ref. \onlinecite{mackenzie96} \\
D &  30, 80 & A.W. Tyler et al., Ref. \onlinecite{tyler98}\\
\noalign{\smallskip}\hline
\noalign{\smallskip}\hline
\end{tabular}
\end{center}
\end{table}
Fits to the optimal doping data are not performed since they yield
unphysical values of the parameters (e.g., values of $1/\tau_0\sim
0$). This is due to the strong increase of the resistivity at optimal
doping and is probably related to the opening of the pseudogap or some
other new physics, which is beyond the scope of our model self-energy.

\begin{figure}[htb] 
\centering 
\includegraphics[width = .27\textwidth, angle=-90]
  {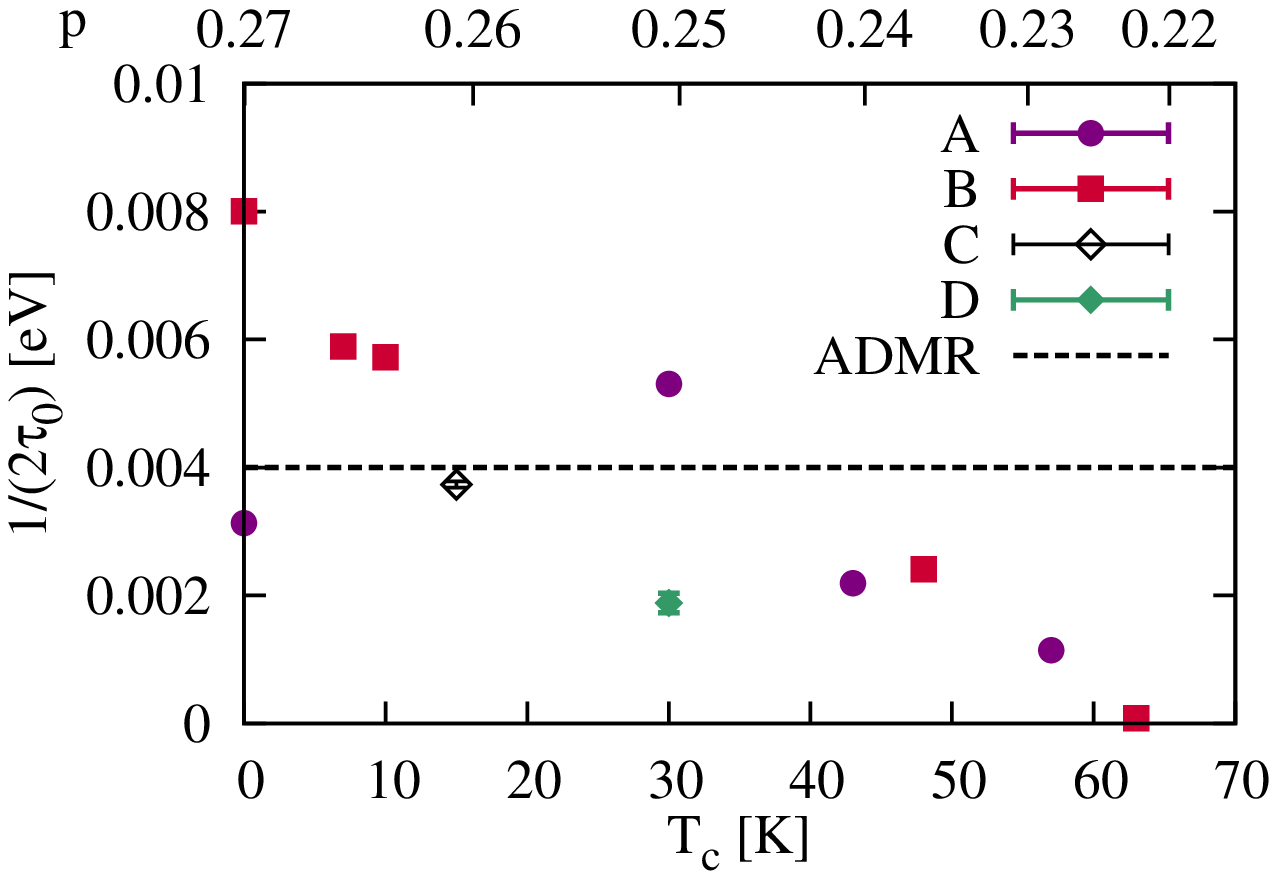}
\includegraphics[width = .25\textwidth, angle=-90]
  {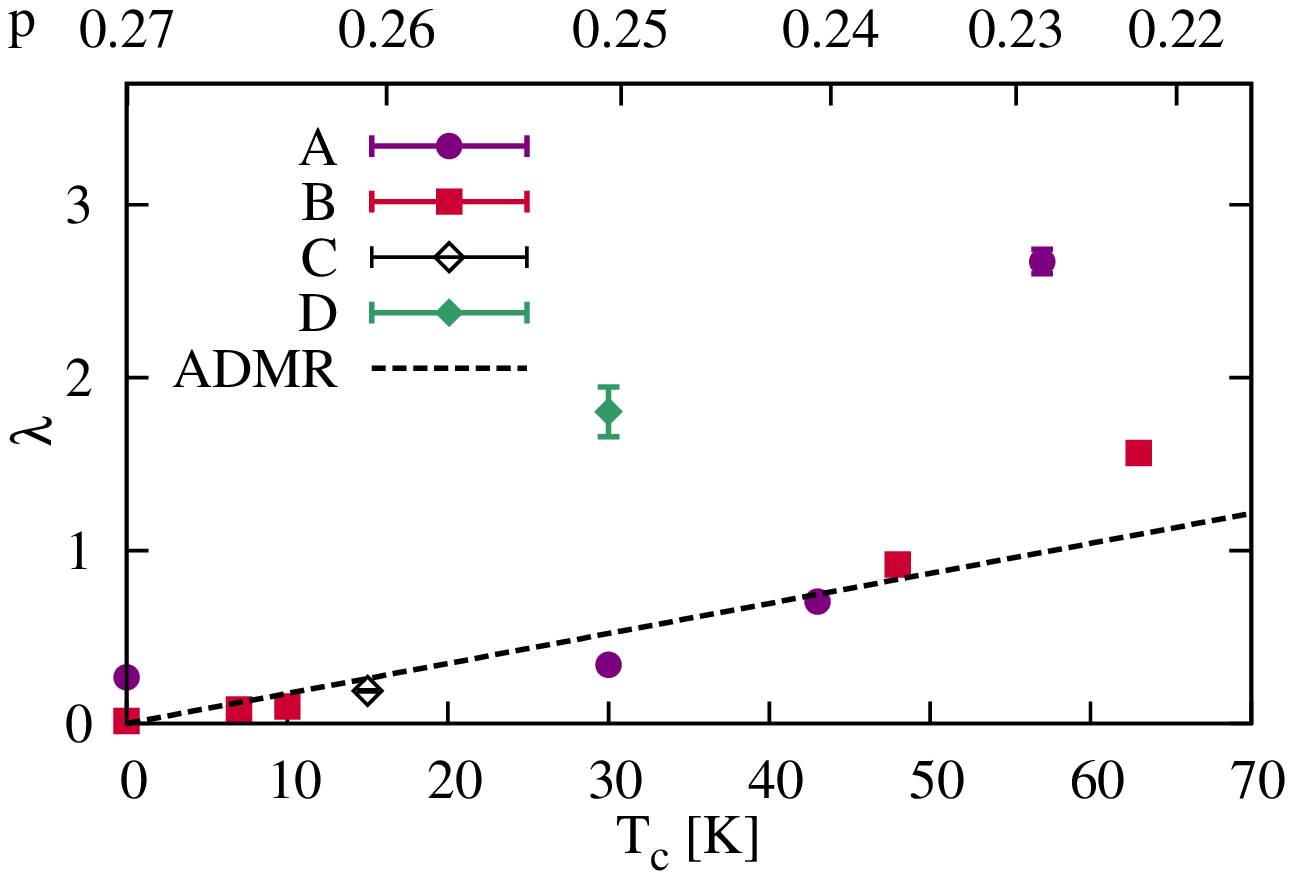}
\includegraphics[width = .26\textwidth, angle=-90]
  {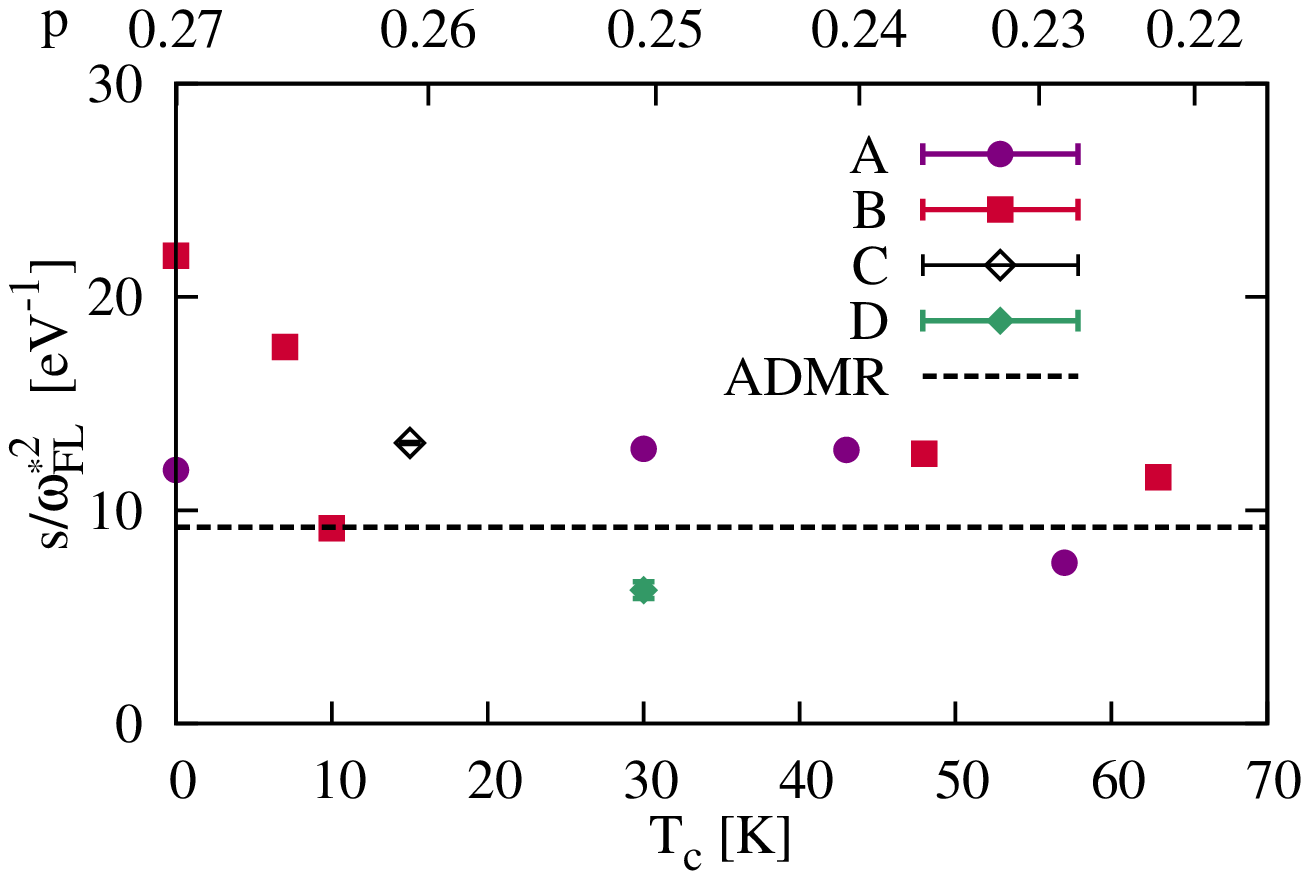}
  \caption{Doping ($p$) or transition temperature ($T_c$) dependence
    of the model parameters extracted from fitting the temperature
    dependence of the resistivity for a range
of Tl2201 samples. The fitted model parameters 
are consistent  with the values and doping dependencies
    extracted from ADMR \cite{kokalj11}. Strength of the impurity
    scattering $1/(2\tau_0)$ shows additional decreasing trend with
    decreasing doping, which might be due to a smaller amount of doped
    interstitial oxygen. The AMFL strength $\lambda$ shows a strong
    increase with decreasing doping in good agreement with the results
    from ADMR \cite{kokalj11}. Good agreement with the self-energy
    model is also found for the doping independent strength of the FL like
    scattering $s/\omega_\textrm{FL}^{*2}$.
 For the doping dependence
    of $T_c$ we use the phenomenological relation (see section
    \ref{sec_modelselfenergy}). 
 $T_c$ values and references for the
    data denoted with A, B, C and D are given in Table \ref{tab1}.  }
\label{fig2}
\end{figure}

The resulting fit parameters together with the ones extracted from
ADMR are shown in Fig. \ref{fig2}.  All parameters are consistent with
the ones extracted from ADMR \cite{kokalj11}. The zero-temperature
scattering rate $1/(2\tau_0)$ seems to show an additional decreasing
trend with increasing $T_c$, which might be attributed to the loss of
interstitial oxygen causing impurity scattering.  The anisotropic
marginal Fermi liquid parameter $\lambda$ increases with $T_c$ as
expected, although it suggests a super-linear increase for $T_c$ close
to the optimal doping.  The parameter $s/\omega_\textrm{FL}^{*2}$ is
slightly larger than extracted from ADMR but still fairly constant
with doping. Similar results were also obtained from the
conductivities of overdoped LSCO \cite{cooper09}.

 Fitting parameters for $T_c>70$ K become unphysical (too
small $1/\tau_0$) and might be a sign of a new physics out of the scope
of our simple model self-energy.

We found that the resulting fit parameter values do not change
significantly if only the zero frequency self-energy is taken into
account, as occurs with the delta function approximation for the
Fermi-function term, Eq. (\ref{eq_partialnf}).

For higher temperatures the measured resistivity shows a linear in $T$
dependence over a broader temperature region than our model
(Fig. \ref{fig1}).  As discussed further below, a smoother saturation
of the self-energy at high $T$ and high $\omega$ may improve the
comparison in this regime.

Saturation of the self-energy may originate in the Mott-Ioffe-Regel
(MIR) limit at which the mean free path
$l=v_{F,0}/(-2\Sigma''_\textrm{max})$
becomes comparable to the
lattice constant and electrons become incoherent. Estimate of
$-\Sigma''_\textrm{max}$ from the MIR limit and our LDA estimate of
$v_{F,0}\sim 1 a$ [eV] gives $-\Sigma''_\textrm{max}\sim 0.5$ eV. This
is in good agreement with our maximal value of the FL part of
self-energy (the main contribution at high $T$) which is $\sim 0.5$
eV. The MIR limit was already successfully applied to the scattering
rate saturation of the optimal and overdoped cuprates
\cite{hussey03a}. It is important to mention, that in the underdoped
regime, the resistivity saturates at a much larger value than expected
from the MIR limit, which may be due to the smaller carrier
concentration \cite{gunnarsson03}.

\subsection{Optical Conductivity}

Experiments do not directly measure the frequency dependent conductivity
but rather the reflectivity or absorption of a thin film or single crystal.
The real and imaginary parts of the conductivity are then extracted
from a Kramers-Kronig analysis \cite{dressel}.
This is only stable and reliable if there is experimental data out
to sufficiently high frequencies.
Furthermore, to aid the physical interpretation of the results
experimentalists often plot the frequency dependent scattering rate
and effective mass that is deduced from an extended Drude model \cite{basov05}.
However, this also requires a knowledge of the plasma frequency
$\omega_p$ (compare Eq. (\ref{eq_plasma})).
As mentioned earlier, the bare band dispersion from LDA
predicts $\omega_p\sim 23000$ cm$^{-1}$ a value which is
larger than extracted from
experiments (15100 cm$^{-1}$ in Ref. \onlinecite{puchkov96}). 

To simplify the analysis and avoid the introduction of new parameters
we compare the results for the model self-energy
directly to the measured reflectivity.
The reflectivity [$R(\omega)$] or absorption [$A(\omega)=1-R(\omega)$] may
be written in terms of the optical conductivity 
\cite{dressel}
\eq{
R(\omega)=
\frac
{1 +\frac{1}{\epsilon_0\omega}|\sigma_{xx}(\omega)|
-\sqrt{\frac{2}{\epsilon_0\omega} \big[ 
       |\sigma_{xx}(\omega)|  -\IM \sigma_{xx}(\omega)     \big]}
 } 
{1 +\frac{1}{\epsilon_0\omega}|\sigma_{xx}(\omega) |
+\sqrt{\frac{2}{\epsilon_0\omega} \big[ 
       |\sigma_{xx}(\omega) | -\IM\sigma_{xx}(\omega)     \big]}
 } ,
\label{eq_romega}
}
in the limit $\IM\sigma_{xx}(\omega) \gg \epsilon_0\omega$, which
is valid in the frequency region of the data.
Here $|\sigma_{xx}(\omega)
|=([\RE\sigma_{xx}(\omega)]^2+[\IM\sigma_{xx}(\omega)]^2)^{1/2}$. 

\begin{figure}[htb] 
\centering 
\includegraphics[width = 0.3\textwidth, angle=-90]
  {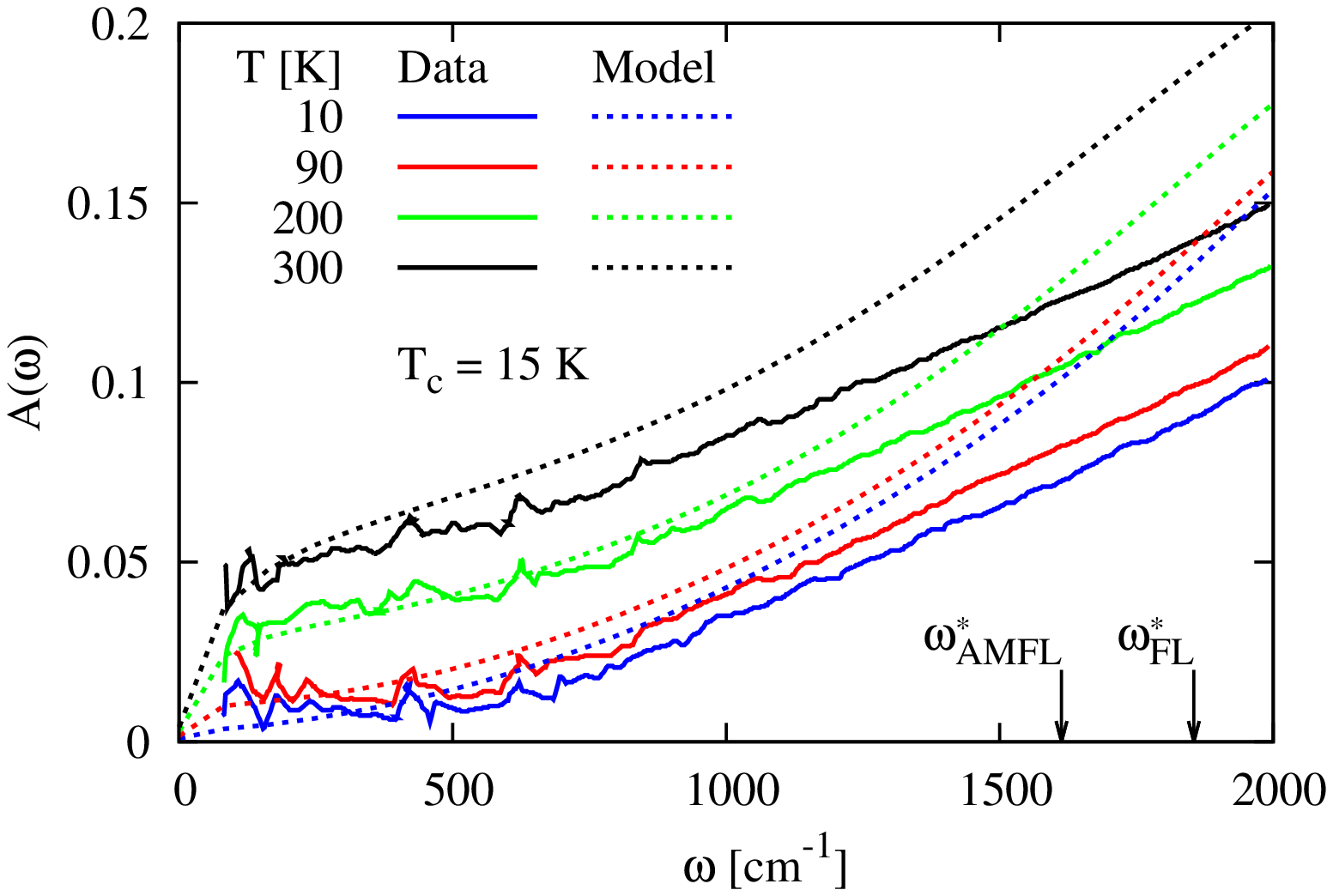}
\includegraphics[width = 0.3\textwidth, angle=-90]
  {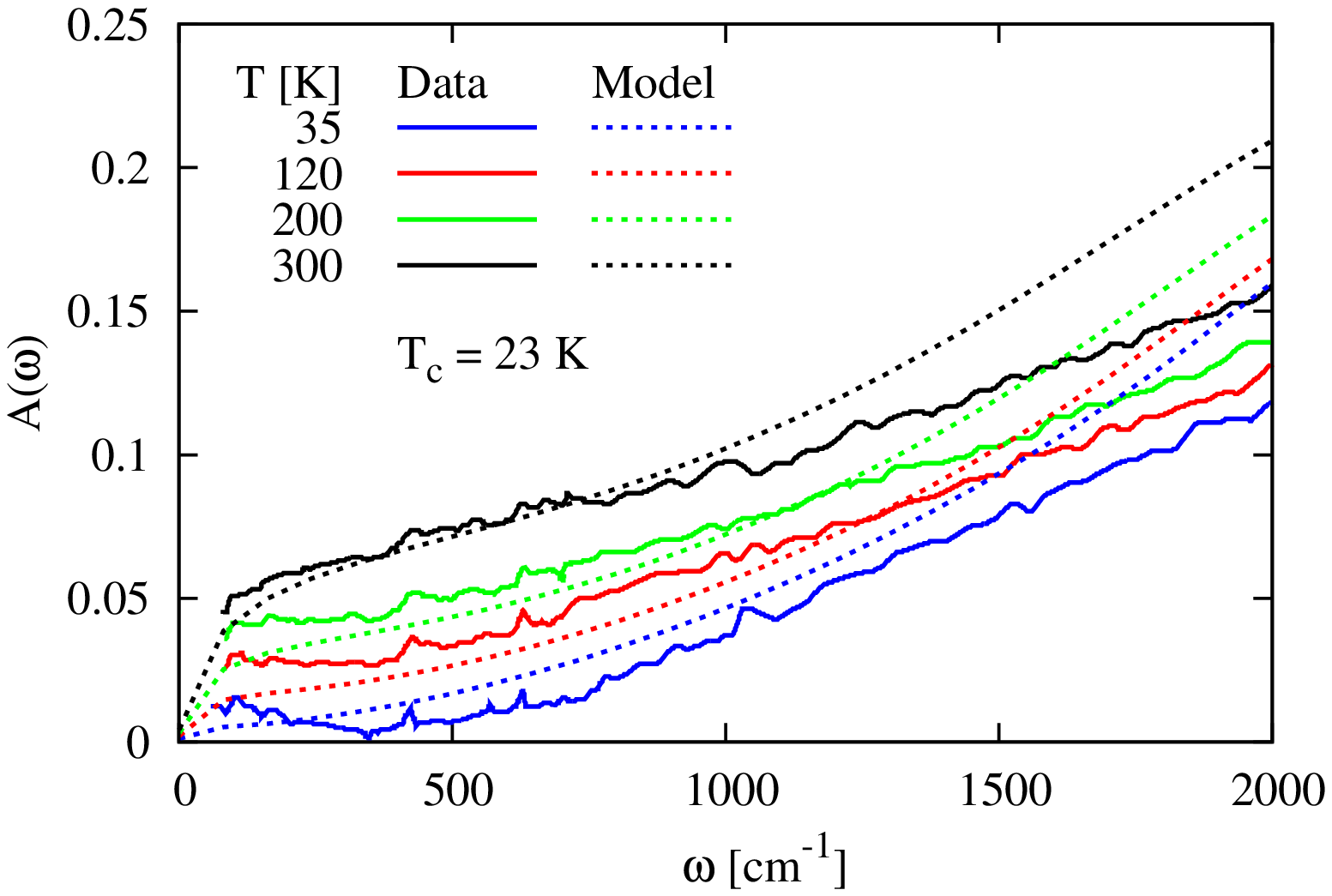}
\includegraphics[width = 0.3\textwidth, angle=-90]
  {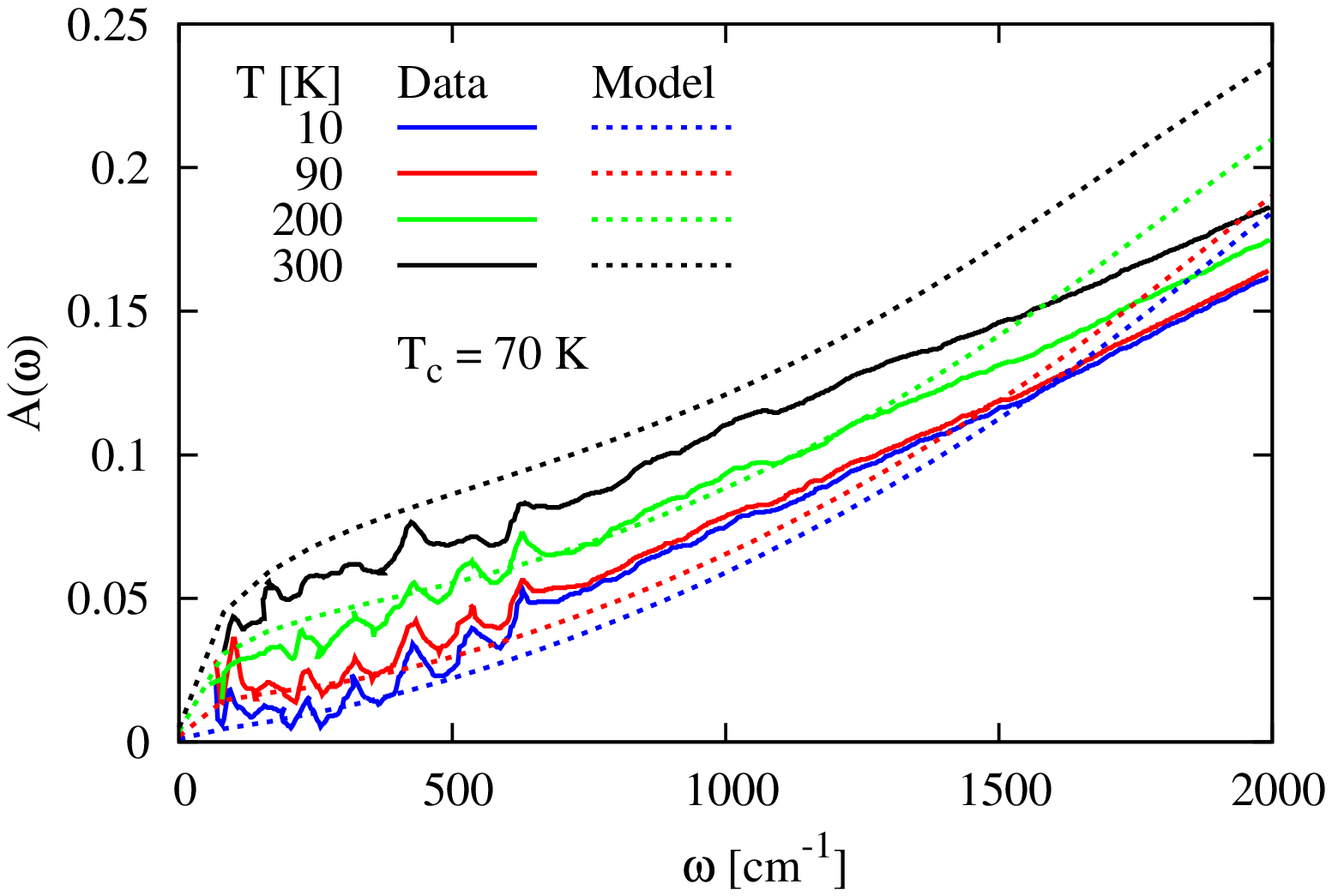}
\caption{ 
Comparison of the measured optical
  absorption spectra $A(\omega)$ for Tl2201 at three different
dopings with the model self-energy.   The latter is
for parameter values
  extracted from  ADMR\cite{kokalj11}. A good description of the
$\omega$, $T$ and $T_c$ dependencies of  $A(\omega)$ is obtained
for $\omega \lesssim 1000$ cm$^{-1}$. For higher $\omega$, $A(\omega)$ shows
a stronger increase with
$\omega$, which could be improved with a softer high-$\omega$ cutoff
for the self-energy.  
This is similar to what is found for the high-$T$  DC 
resistivity in Fig. \ref{fig1}. 
Data for $T_c=$ 15 K and 70 K are from Ref. \onlinecite{ma06}
and data for $T_c=23$ K are from Ref. \onlinecite{puchkov96}. 
 }
\label{fig3}
\end{figure}


Comparison of our results, obtained with Eqs. (\ref{eq_romega}) and
(\ref{eq_sigmaxx}) and model 
self-energy parameters extracted from ADMR \cite{kokalj11}, with the 
measured absorption is shown in Fig. \ref{fig3}.  Agreement at low
frequencies ($\omega \lesssim 1000$ cm$^{-1}$) is quite satisfactory.
We consider this is quite impressive given that no additional fitting
parameters beyond those extracted from ADMR\cite{kokalj11} have been
introduced. 

At higher frequencies our model self-energy predicts an absorption
that is too large compared to experimental data. This discrepancy could be
fixed by incorporating a smoother high-frequency saturation making
the self-energy more slowly increasing with $\omega$ and rounding its
behaviour at the high frequency cutoff ($\omega_\textrm{AMFL}^*\simeq
\omega_\textrm{FL}^*\simeq 1700$ cm$^{-1}$, see Fig. \ref{fig3}). One
way of smoothing the high $T$ and $\omega$ behavior could be in
adopting the phenomenological
approach of Refs. \onlinecite{hussey03a, hussey06}, where
the saturation of the scattering rate is applied by the
"parallel-resistor" formula,  which means the imaginary part of
the self-energy (\ref{eq_selfenergy}) 
is replaced according to
\eq{
  \frac{1}{\Sigma''_\textrm{eff}(\phi,T,\omega)}
=
  \frac{1}{\Sigma''_\textrm{ideal}(\phi,T,\omega)}
+ \frac{1}{\Sigma''_\textrm{max}}.
 } 
 Here $\Sigma''_\textrm{ideal}(\phi,T,\omega)$ is the self-energy
 without high-frequency cutoffs and
 $\Sigma''_\textrm{max}$ is the maximal or
 saturated value of the imaginary part of self-energy,
and can be treated as a free parameter.
In the MIR picture described above this parameter is
 estimated to have a value $\sim \hbar v_F/a$, where
$a$ is the lattice constant.

Using two different model self-energies Norman and Chubukov
\cite{norman06} performed a detailed analysis of the frequency
dependent conductivity for optimally doped
Bi$_2$Sr$_2$Ca$_{0.92}$Y$_{0.08}$Cu$_2$O$_{8+\delta}$.  They deduced a
flattening of the frequency dependence of the scattering rate near a
cutoff energy of order 0.3 eV.
The high-frequency cutoff may also be observed in ARPES as a kink or
''waterfall'' in the QP dispersion due to a noticeable change in 
$\partial \Sigma'/\partial \omega$, particularly
if it obtains a  value $\gtrsim 1$  
(for example see Refs. \onlinecite{chang08,zhu08}).   

The cutoffs give some insight into the underlying physics since they
tell us the energy scales of the excitations (e.g., spin fluctuations,
particle-hole excitations) which the electrons are scattering off
\cite{norman06}. On the other hand, cutoffs may also reflect the limiting
scattering rate (e.g., given by the sum rule
Eq. (\ref{eq_sigmasumrule}) or the entry into the MIR limit
\cite{hussey03a, gunnarsson03}) or  entrance into the incoherent regime.

\section{Hall effect}
\label{sec_Halleffect}
\subsection{Hall coefficient}

The Hall coefficient in the weak field limit is given by
\eq{
R_H=\frac{\sigma_{xy}^{(1)}}{B_z [\RE\sigma_{xx}]^2},
}
where $B_z$ is the magnetic field in the $z$ or $c$ direction,
$\sigma_{xy}^{(1)}$ is the Hall conductivity proportional to $B_z$, and
$\RE \ \sigma_{xx}$ is the in-plane DC conductivity (see
Sec. \ref{sec_dcconductivity}).  

A diagrammatic calculation of the Hall conductivity is given in
Ref. \onlinecite{kohno88}, 
leading to
\eq{
\sigma_{xy}^{(1)} = \frac{-\im eB_z}{2}\sum_k 
\int \frac{d \omega}{2\pi}
(- \frac{\partial n_F(\omega)}{\partial \omega}) [J_x \tilde \partial_y  J_y] [G^R
\tilde \partial _x G^A] , 
\label{eq_sigmaxy1}
}
 where $[A\tilde \partial_\mu B]= A\partial_{k_\mu} B
 - (\partial_{k_\mu} A) B $, $J_\mu$ is a current vertex which we
 approximate with $-e v_{0,\mu}$ by neglecting the vertex corrections,
 and $G^{R(A)}$ is the retarded (advanced) Green's functions.

The expression in Eq. (\ref{eq_sigmaxy1}) for the Hall conductivity can be further
simplified with the following approximations. First, we neglect the
term $\partial_{k_x} \Sigma'(k,\omega)$, which arises from 
differentiation of the Green's function and is present also as a first
trivial correction to the vertex, which we also neglect.
We find that calculations with this correction do not 
significantly change the results,
because our $\Sigma'(k,\omega)$ is odd-in-$\omega$.
  Then we linearize the dispersion in the $k_r$ direction
around the Fermi surface, Eq. (\ref{eq_epsilon0k}), 
and approximate the integral over $k_r$, as was
similarly done for the DC conductivity
(Sec. \ref{sec_dcconductivity}). Using the symmetry 
$\sigma_{xy}^{(1)}=-\sigma_{xy}^{(1)}$ and manipulations similar to 
those of Ong in Ref.  \onlinecite{ong91} leads to 
\eqa{
\sigma_{xy}^{(1)} =
\frac{e^3 B_z}{4\pi^2 d}
\int d \phi
[-{\bf v}_{0,F}(\phi)\times \partial_\phi{\bf v}_{0,F}(\phi) ]_z\nonumber\\
\times \int d\omega (- \frac{\partial n_F(\omega)}{\partial \omega})
\frac{1}{(-2\Sigma''(\phi,\omega))^2 },
\label{eq_sigmaxy1_summary}
}
where we have also used that our self-energy depends only on $\phi$ in
momentum space. A more detailed derivation can be found in
Appendix \ref{sec_detailsonderivation}. 

\subsection{Comparison with the Boltzmann equation}

Ong has given an elegant geometrical interpretation of the Hall
conductivity $\sigma_{xy}$ for a two-dimensional Fermi liquid\cite{ong91}. 
It is proportional to the area swept out by the scattering length 
or mean-free path
${\bf l}(\phi) \equiv {\bf v}_F(\phi) \tau(\phi) $ 
as one traverses the Fermi surface.
This illustrates how the Hall effect is sensitive to anisotropy in
the Fermi surface via the Fermi velocity ${\bf v}_F(\phi)$ and
and to anisotropy in the scattering time $\tau(\phi)$. 

Eq. (\ref{eq_sigmaxy1_summary}) is consistent with the expression derived
from the Boltzmann equation \cite{hussey03a} and with Ong's
geometric expression\cite{ong91}. If
the frequency dependence of the self-energy close to $\omega=0$ is neglected,
then   
\eq{
\int d\omega (- \frac{\partial n_F(\omega)}{\partial \omega})
\frac{1}{(-2\Sigma''(\phi,\omega))^2}
\simeq
\frac{1}{(-2 \Sigma''(\phi,0))^2}.
\label{eq_int_omega}
}

To make the comparison with the Boltzmann equation and relaxation time
approximation more explicit, we start with the Boltzmann equation result for
the
Hall conductivity \cite{hussey03a}, 
\eq{
  \sigma_{xy}^{(1)}=\frac{e^3}{2\pi^2 d } \int d^2k (-\frac{\partial
    n_F(E_k)}{\partial E_k}) \frac{v_x}{\Gamma} {\bf v}\times {\bf B}
  \cdot {\bf \nabla}(\frac{v_y}{\Gamma}).
  }
 Here $E_k$ is the quasi-particle (QP)
dispersion, ${\bf v}=(v_x,v_y)$ is the QP velocity, and $\Gamma$ is
the scattering rate, which are all $k$-dependent. The integral goes over
the first Brillouin zone in two dimensions.
  Symmetrizing the expression with the use of
$\sigma^{(1)}_{xy}=-\sigma^{(1)}_{yx}$ and applying 
\eq{
  \frac{v_x}{\Gamma} {\bf v}\times{\bf B} \cdot {\bf
    \nabla}(\frac{v_y}{\Gamma}) = {\bf v}\times{\bf B}\cdot
  (\frac{1}{\Gamma} {\bf \nabla}(\frac{v_x
    v_y}{\Gamma})-\frac{v_y}{\Gamma^2} {\bf \nabla}(v_x)),
 }
 leads to
the following expression, 
\eqa{ \sigma_{xy}^{(1)}=\frac{e^3}{4\pi^2 d
  } \int d^2k
  (-\frac{\partial n_F(E_k)}{\partial E_k})  {\bf v}\times{\bf B} \nonumber\\
 \cdot (v_x {\bf \nabla}(v_y) -v_y {\bf
    \nabla}(v_x)) \frac{1}{\Gamma^2}.  
}
 Furthermore, if the integral
over the 
2D Brillouin zone is decomposed into the integrals over $\phi$ and $k_r$
and in addition the QP dispersion is linearized close to the Fermi surface with
$E_k\simeq v_{F,r}(\phi) (k_F(\phi)-k_r)$ and $\Gamma\simeq
\Gamma(\phi)$, then the integral over $k_r$ may be performed
(neglecting band edge effects) and we are left only with the integral
over $\phi$. For the magnetic field in the $z$ direction one can then
rewrite $\sigma_{xy}^{(1)}$ in a similar form as in
Eq. (\ref{eq_sigmaxy1_summary}) if the integral over $\omega$ in Eq.
(\ref{eq_sigmaxy1_summary}) is replaced with
$\frac{1}{\Gamma(\phi)^2}$
(similar to Eq. (\ref{eq_int_omega})).
We should note here that the expression
derived from the Boltzmann equation includes only renormalized
quasi-particle 
entities, while Eq. (\ref{eq_sigmaxy1_summary}) includes only
non-renormalized quantities. This is not a problem since the
renormalization cancels by taking $v_F=Z v_{0,F}$ and $\Gamma=-2 Z
\Sigma''$. However, this might not be the case, if the shape of the
non-interacting Fermi surface is changed due to the renormalization. This does
not happen for our model self-energy, since its real part is always
zero at $\omega=0$ due to 
the imaginary part being an even function of frequency.

The relationship to Ong's geometric interpretation is more straight
forward. If the integral over $\omega$ in
Eq. (\ref{eq_sigmaxy1_summary}) is approximated as in
Eq. (\ref{eq_int_omega}), one can write
\eqa{
\sigma_{xy}^{(1)} &=&
\frac{-e^3 B_z}{4\pi^2 d}
\int d \phi
[{\bf l}(\phi)
\times \partial_\phi
{\bf l}(\phi)]_z,
\label{eq_sigmaxx1_withl}
}
where 
\eq{
{\bf l(\phi)}= {\bf v}_{0,F}(\phi)/[-2\Sigma''(\phi,0)]
\label{eq_lphi}
} 
is the
mean free path used in the Ong's geometrical interpretation of the
Hall conductivity. From this expression it is nicely seen that the
renormalization $Z$ cancels and that the Hall conductivity is
determined by the mean free path on the Fermi surface.

\subsection{Comparison with experiment}
\label{sec_Comparisonwithexperiment}

The zero temperature ($T=0$) value of the Hall coefficient $R_H$ for a
circular Fermi surface corresponds to $1/(e n_e)$ with $n_e$ being the
density of electrons. Deviations from this value depend on the shape
of the Fermi surface. If for our tight-binding band structure we
assume a rigid band shift from the highly overdoped to optimally doped
regime, then the $T=0$ value of $R_H$ is expected to change by less
than 10\%.
Temperature broadening affects $\RE\sigma_{xx}$ and
$\sigma_{xy}^{(1)}$ only at higher $T$ and is within our model
estimated to result in a relative change of $\sim 10\%$ at
$T\sim 200$ K. The $T$ broadening effect is reduced in $R_H$ and is
estimated to be $\lesssim 0.5\times 10^{-10} $ m$^3/$C.
In contrast to the above relatively small changes with temperature and
doping, experiment shows that $R_H(T)$ can vary by as much as 100\%,
as $T_c$ varies from 0 to 50 K in the overdoped regime (compare 
Figure \ref{fig6}).

\begin{figure}[htb] 
\centering 
\hspace{-15mm}
\includegraphics[width = 0.2\textwidth, angle=-90]
{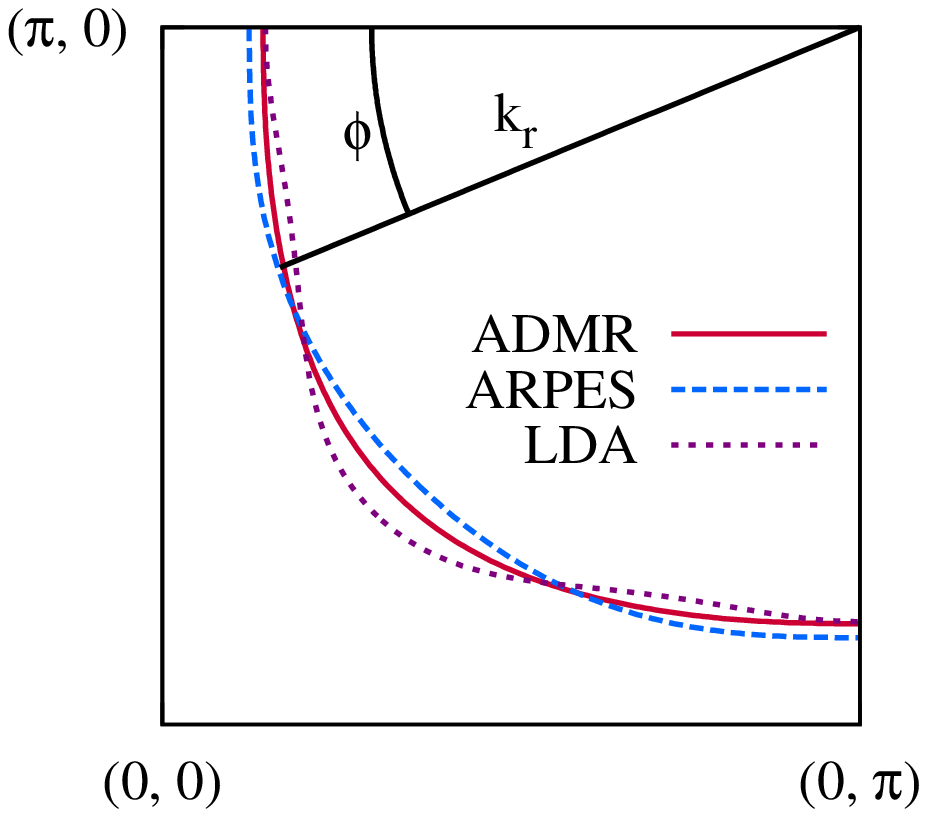} \hspace{-13mm}
\includegraphics[width = 0.23\textwidth, angle=-90]
{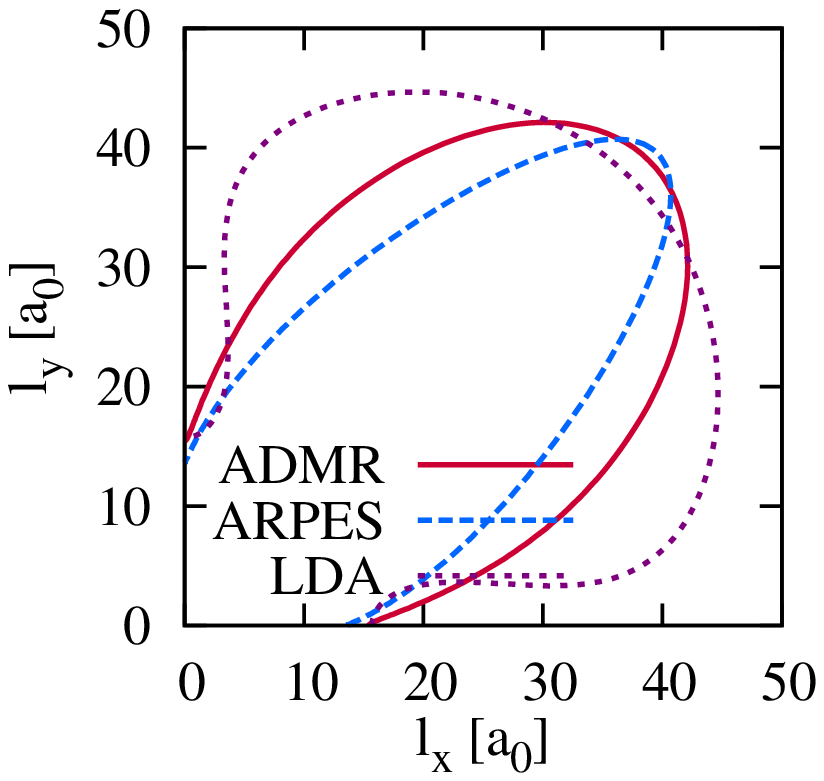}\hspace{-15mm}
\caption {(Left) Three different approximations for the shape of the
  Fermi surface (FS). The FS deduced from ADMR measurements
  \cite{abdel07}, ARPES measurements \cite{plate05}, and tight-binding
  approximation to the LDA calculations \cite{peets07,kokalj11} are
  denoted with ADMR, ARPES, and LDA, respectively. (Right) Curves of
  the mean-free path ${\bf l}(\phi)$ as one goes around the FS for the
  three different FSs. According to Ong's geometric interpretation
  \cite{ong91} the encircled area is proportional to
  $\sigma_{xy}^{(1)}$ [Eq. (\ref{eq_sigmaxx1_withl})].  Although the
  shapes of the FSs do not change much between different
  approximations, the mean-free paths and the encircled areas in ${\bf
    l}$ space change substantially. The main difference comes from the
  curvature of the FS ($\partial_\phi {\bf v}_{0,F}(\phi)$) close to
  $\phi \sim \pi/8$. The absolute value of $R_H$ is therefore very
  sensitive to the shape of the FS. The mean-free path ${\bf l(\phi)}$
  was calculated with our model self-energy for $T=100$ K, and
  $T_c=30$ K.  }
\label{fig4}
\end{figure}

\begin{figure}[htb] 
\centering 
\includegraphics[width = 0.3\textwidth, angle=-90]
{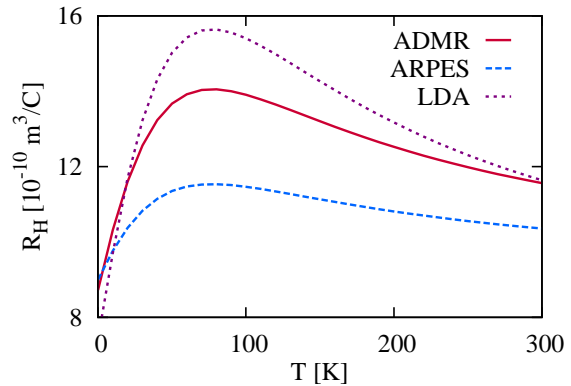}
\caption{Calculated temperature dependence of the
Hall coefficient ($R_H$) for
three slightly  different Fermi surfaces (see
  Fig. \ref{fig4}). The absolute change or maximal value of $R_H$ depends
 strongly on the shape of the Fermi surface. All
curves are calculated with the same scattering rate,
using our self-energy model for
  $T_c=30$ K. Temperature broadening effects
due to the Fermi-Dirac distribution are not taken into account since
  they are small ($<10$\%). The slightly different
values of $R_H$ at $T=0$, where the scattering
  is dominated by impurities and is therefore isotropic, also comes from
  small changes in the Fermi surface shape.   
}
\label{fig5}
\end{figure}

\begin{figure}[htb] 
\centering 
\includegraphics[width = 0.3\textwidth, angle=-90]
  {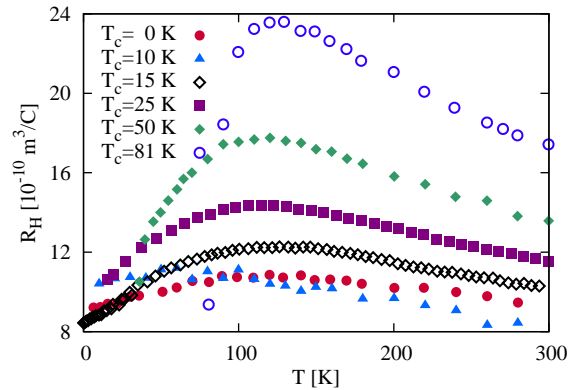}\\
  \caption{ Temperature dependence of the measured Hall coefficient $R_H$
    for several different $T_c$s. These all correspond to dopings
for which there is a large hole Fermi surface
    and show a non-monotonic temperature dependence that                  
    increases with increasing $T_c$. For $T \lesssim T_c$ $R_H$ may
    be strongly suppressed by the superconducting
 transition.  Data for $T_c=$ 0 K and
    81 K are for polycrystalline samples measured in
    Ref. \onlinecite{kubo91}, data for $T_c=$ 10 K and 50 K are from
    Ref. \onlinecite{manako92}, $T_c=$ 15 K data is from
    Ref. \onlinecite{mackenzie96}, and $T_c=$ 25 K data is from
    Ref. \onlinecite{hussey96}.  }
\label{fig6}
\end{figure}

As can be seen from Eq. (\ref{eq_sigmaxy1_summary}) for
$\sigma_{xy}^{(1)}$ and Eq. (\ref{eq_dc_cond}) for $\RE\sigma_{xx}$
the temperature dependence of the Hall 
coefficient comes from the $T$-dependence of the anisotropy of the
scattering rate. This becomes
more apparent if we rewrite the Hall coefficient in the following
form
\eq{
R_H=
 \frac{
\int d\phi f_H(\phi) \frac{1}{(-2\Sigma''(\phi))^2} 
}
{
[\int d\phi f_{DC}(\phi) \frac{1}{(-2\Sigma''(\phi))}]^2
},
\label{eq_rhdphi}
} 
where we have neglected the $T$-broadening effects. $f_H(\phi)$ is
the $\phi$ dependent coefficient (corresponding to the Hall
conductivity), which needs to be integrated over $\phi$ and depends
only on bare-band dispersion. $f_{DC}(\phi)$ is similar to
$f_H(\phi)$, but for  the DC conductivity (see Appendix
\ref{sec_fhphi}). The only $T$-dependent quantity in the above
equation is the self-energy and its $T$-dependent anisotropy is
responsible for $T$-dependent $R_H$. This is in agreement with
results in Ref. \onlinecite{stojkovic97}. 
However, the absolute change of $R_H$ with temperature depends strongly on the
shape of the Fermi surface. This is demonstrated in  Figs. \ref{fig4} and \ref{fig5}. 

\begin{figure}[htb] 
\centering 
\includegraphics[width = 0.3\textwidth, angle=-90]
  {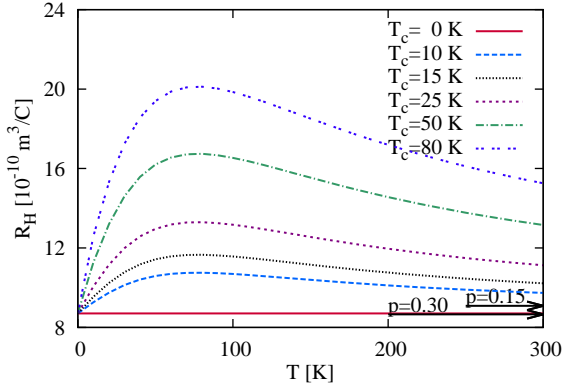}\\
  \caption{Temperature
dependence of the  Hall coefficient $R_H$ calculated with our model self-energy for several
    $T_c$s and for the ADMR Fermi surface \cite{abdel07}. Results should be compared
    with Fig. \ref{fig6}. The figure illustrates how
decreasing doping (increasing $T_c$
    or $\lambda$ or anisotropy of the self-energy) leads to a large change in
  the   magnitude of the temperature dependence. 
All results are obtained for the same Fermi surface
 and only the anisotropy of the self-energy is changed. Arrows
shown in the lower right indicate the weak dependence
on the band structure change with doping,
    indicating the absolute shift of the zero temperature
value of $R_H$ ($T_c=0$) for two different
    dopings ($p=0.15$ and $p=0.30$) as given by a rigid band shift.  }
\label{fig7}
\end{figure}

The overall doping (or $T_c$) and $T$ dependence of 
the measured and calculated
 $R_H$  are
shown in Fig. \ref{fig6}  and \ref{fig7}, respectively. 
The temperature  dependence of $R_H$ suggests, that the scattering anisotropy
strongly (linearly) increases at low $T$ (in our model due to the AMFL
part of self-energy), reaches its maximum at $\sim 110$ K and then the
scattering slowly becomes more isotropic again as the  FL part of 
the self-energy model begins to dominate. 

The fact that for $T_c=0$ the experimental $R_H$ shows
a small $T$-dependence (see
Fig. \ref{fig6}) represents a problem for our model, since the 
model has no anisotropy for $T_c=0$.
However, there was no ADMR data for $T_c=0$ and so it is
possible that the anisotropy actually does not go to 0 as $T_c \to 0$,
or perhaps that our assumption that the $T^2$ term is strictly isotropic
needs to be relaxed.

Comparison of our results (Fig. \ref{fig7}) with the measured $R_H$
(Fig. \ref{fig6}) shows qualitative, and to some extent also
quantitative, agreement. However, our $R_H$ does reach a maximum for
$T\sim 80$ K, while the maximum appears at higher T ($\sim 110$ K) in
experiment (Fig. \ref{fig6}). In order to get a better comparison the
FL part of our self-energy model should be reduced (smaller
$s/\omega_\textrm{FL}^{*2}$).  Also inclusion of a smoother high
 frequency cutoff could move the maximum in our $R_H(T)$ to higher
 $T$. 

In fitting our model to $R_H$ it turns out that one parameter is free
(one of $1/(2\tau_0)$, $\lambda$ or $s/\omega_\textrm{FL}^*$).
This is because 
the absolute value of $R_H$ is unchanged by a re-scaling of the scattering
time, as can be seen from Eq. (\ref{eq_rhdphi}).
This    is 
closely related to $R_H$ not depending on $\tau$ in a simple FL picture.

\subsection{Hall angle}

\begin{figure}[htb] 
\centering 
\includegraphics[width = 0.3\textwidth, angle=-90]
  {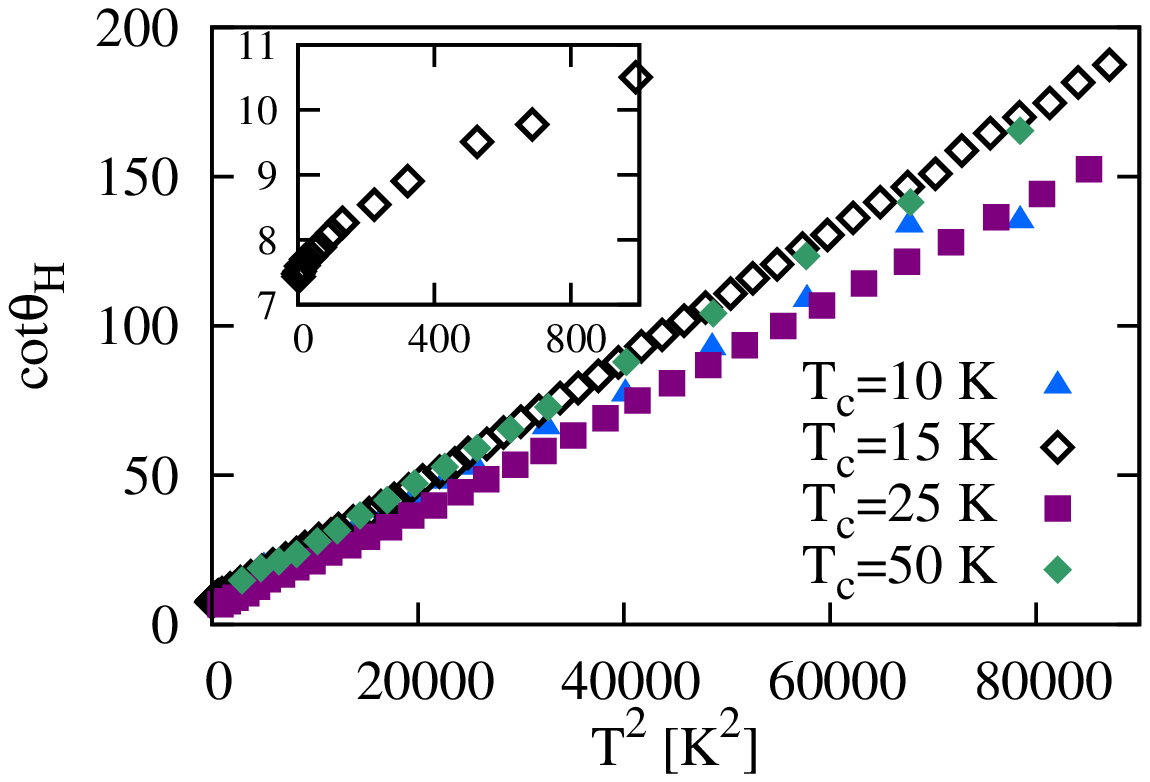}\\
\includegraphics[width = 0.3\textwidth, angle=-90]
  {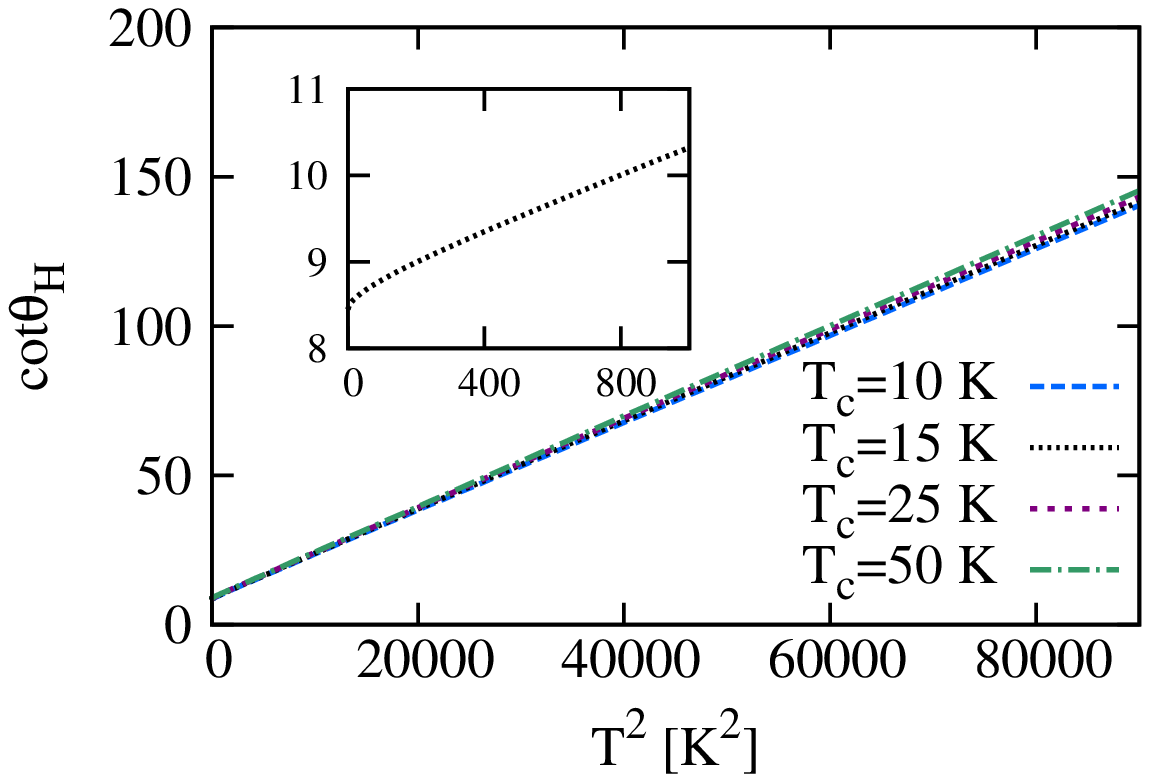}\\
  \caption{ Temperature dependence of the Hall angle $\cot
    \theta_H$. (Top) Experimental data show
that $\cot \theta_H$ has a linear
    dependence on $T^2$ and a weak doping dependence. 
A similar dependence is found for our model (Bottom), 
where the linear dependence of $\cot
    \theta_H$ on $T^2$ comes predominantly 
from the Fermi liquid like scattering
    ($\Sigma''_\textrm{FL}$)
that  gives the scattering rate on the nodal part of the Fermi surface 
(which also gives the dominant contribution to the Hall
    conductivity, see Fig. \ref{fig4}). 
(Top) The Inset shows a small down-turned
    deviation from the $T^2$ dependence of $\cot \theta_H$ at low $T$ (for
    $T_c=15$ K). This is also obtained within our model (bottom
    inset), although not as pronounced as in the experimental data. 
Experimental data for $T_c=$ 10 K and 50 K are 
from Ref. \onlinecite{manako92}. $T_c=$ 15 K
    data is from Ref. \onlinecite{mackenzie96} and $T_c=$ 25 K data is
    from Ref. \onlinecite{hussey96}. Our model results are calculated
    with the ADMR Fermi surface.}
\label{fig8}
\end{figure}

The Hall angle is defined by \eq{ 
\cot \theta_H(T) \equiv
  \frac{\RE\sigma_{xx}}{\sigma_{xy}^{(1)}}= \frac{\rho_{xx}(T)}{R_H(T)
    B_z}.
  }
 Since our model can describe the temperature dependence
of the intra-plane resistivity and Hall coefficient, as we showed
above, it must also describe the Hall angle.  Here, we examine the
temperature dependence of $\cot\theta_H$ in order to point out that
the observed $T^2$ dependence of $\cot\theta_H$ (cf. Figure
\ref{fig8}) naturally follows from our model self-energy and that
there is therefore no need to evoke more exotic theories in order to
explain the qualitatively distinct temperature dependence of
$\rho_{xx}$ and $\cot\theta_H$.

Experimental data and our results are shown in Fig. \ref{fig8} and
provide additional support for our model self-energy. In particular,
the linear dependence of $\cot \theta_H$ on $T^2$ supports the $T^2$
dependence of the isotropic part of self-energy or scattering rate in
the nodal direction. That is because $\cot \theta_H$ is dominated by
the isotropic part ($\Sigma''_\textrm{FL}$), while it suppresses the
anisotropic part ($\Sigma''_\textrm{AMFL}$) of the self-energy. This
point was previously emphasized by Carrington et
al.,\cite{carrington92} Ioffe and Millis \cite{ioffe98} and by
Stojkovic and Pines \cite{stojkovic97} (see also Ref.
\onlinecite{fruchter07}). To show this more explicitly, we use a
similar expression to the one in Eq. (\ref{eq_rhdphi}), approximate
$f_H(\phi)$ and $f_{DC}(\phi)$ with a constant, and perform the
integrals over $\phi$. This leads to
\eq{
\cot \theta_H \propto
-\Sigma_\textrm{FL}''(0)
[1+
\frac{\Sigma_\textrm{FL}''(0)}{\Sigma_\textrm{FL}''(0)+\Sigma_\textrm{AMFL}''(0,0)}
]^{-1}.
}   
It turns out that the temperature dependence
of $\cot \theta_H$ is dominated by the
first factor, because the second factor is weakly temperature
dependent. 
For more details see Appendix \ref{sec_effectofanisotropy}.  
Hence, the Hall angle is dominated by isotropic
scattering or by the region on the Fermi surface with the weakest
scattering or the longest mean-free-path, while the effect of
anisotropic scattering is suppressed.  
Further suppression of anisotropic part comes 
from the anisotropy of $f_H(\phi)$, which is larger in the nodal and smaller
in the antinodal direction. 

Although the effect of $\Sigma_\textrm{AMFL}''$ on $\cot \theta_H$ is
small (note $T^2$ dependence in Fig. \ref{fig8}), it still
changes the pure $T^2$ dependence of $\cot \theta_H$ to $T^n$ with
$n\lesssim2$. Values of $n<2$ were actually observed in 
YBCO (Ref. \onlinecite{Castro04}) and Bi2201
(Ref. \onlinecite{ando99, fruchter07}) where $n$ changes from $\sim 1.8$ in 
the optimal or underdoped regime to $n\sim1.6$ in the  overdoped regime. Our
model predicts $n=2$ in the highly overdoped regime where the AMFL part
of the self-energy is zero, but could predict $n<2$, if the smoother
high-frequency cutoff were introduced. This would make the $T^2$
dependence of FL like self-energy more linear in $T$ for higher $T$,
 observed experimentally in Bi2201
(Ref.  \onlinecite{hussey03a}). 
However, with decreasing doping and consequently increasing
anisotropy our model would predict a further decrease of $n$, which is the
opposite trend to that observed experimentally
Bi2201\cite{ando99,fruchter07}.
 In contrast, a different model with strong anisotropic impurity scattering,
 an anisotropic term $\propto T^2$, and  a smooth
   high-frequency saturation yields an increase of $n$ with increasing
   anisotropy\cite{hussey03a}.  
For Tl2201 no change of $n$ with
doping was observed so far, which might be due to a more square-like
Fermi surface and therefore the decreased effect of anisotropy on $\cot
\theta_H$.   

Our anisotropic self-energy model is therefore capable of
simultaneously describing the linear in $T$ part of the DC
conductivity and $T^2$ dependence of $\cot \theta_H$ over a wide
doping range from optimal to the heavily overdoped region. This shows, that
there is no need to introduce more exotic theories with two types of
quasi-particles (e.g., spinons and holons) with different scattering 
rates \cite{anderson91,coleman96}, to capture the
qualitatively different temperature dependence of
$\rho_{xx}$ and $\cot \theta_H$.

\section{Intra-layer magnetoresistance}
\label{sec_Intralayermagnetoresistance}

In this section we consider the intra-layer magnetoresistance which is 
$\propto B_z^2$ for weak magnetic fields in $z$ direction
($B_z$).  Within the Boltzmann theory the 
corresponding intra-layer conductivity is
$\sigma_{xx}=\sigma_{xx}^{(0)}+\sigma_{xx}^{(2)}$ where
$\sigma_{xx}^{(0)}$ is the part of the conductivity independent of
magnetic field, which is given by 
 (compare Eq. (\ref{eq_dc_cond})),
\eq{
\sigma_{xx}^{(0)}= \frac{ e^2}{4 \pi^2 d}  
 \int d\phi
\frac{k_F(\phi) }{\cos \theta} |{\bf l}(\phi)|,
}
while $\sigma_{xx}^{(2)}$ is given by \cite{hussey03a, zheleznyak99}
\eq{
\sigma_{xx}^{(2)}=- \frac{e^4 B_z^2}{4\pi^2 d}\int d\phi
\frac{\cos\theta}{k_F(\phi)} l(\phi) |\partial_\phi {\bf l}(\phi)|^2.
\label{eq_sigmaxx2}
}
${\bf l}(\phi)$ is the mean free path on the Fermi surface at angle
$\phi$ (see Eq. (\ref{eq_lphi}) and Fig. \ref{fig4}), while $\theta$ is an angle between the
Fermi surface direction and the direction ${\bf e}_\phi$ (perpendicular to
$k_r$), which also depends on $\phi$.  The change of the intra-layer
resistivity $\Delta \rho_{xx}^{(2)}$ due to the magnetic field is
obtained with the inversion of the conductivity tensor.
 \eq{
 \frac{\Delta
    \rho_{xx}^{(2)}}{\rho_{xx}} = -\frac{\sigma_{xx}^{(2)}}
  {\sigma_{xx}^{(0)}} - \bigg ( \frac{\sigma_{xy}^{(1)}}{\sigma_{xx}^{(0)}}
  \bigg ) ^2.  
}
For reasons of simplicity we use Boltzmann results for conductivities
($\sigma_{xx}^{(0)}$, $\sigma_{xy}^{(1)}$ and $\sigma_{xx}^{(2)}$), which
can all be expressed with integrals over $\phi$ of different
expressions involving ${\bf l}(\phi)$ (see also Ref.
\onlinecite{zheleznyak99}). No temperature broadening effect is taken into
account, which was found to be small for the Hall effect
(Section \ref{sec_Comparisonwithexperiment}).

\begin{figure}[htb] 
\centering 
\includegraphics[width = 0.3\textwidth, angle=-90]
  {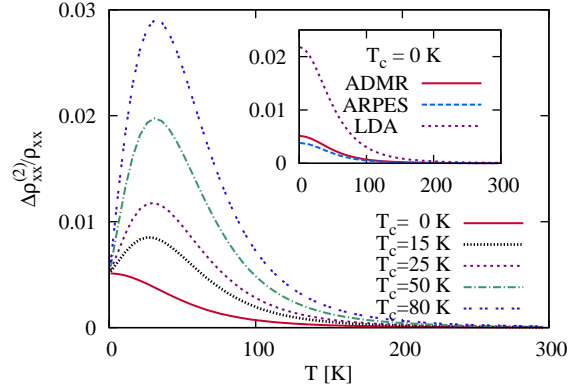}\\
  \caption{ Temperature dependence of the intra-layer
  magnetoresistance $\Delta \rho_{xx}^{(2)}/\rho_{xx}$ for various
  $T_c$ (or strength of anisotropic scattering) 
  calculated with the AMFL model for the ADMR Fermi surface and for $B_z=10$
  T. Temperature dependence of the result for $T_c =0$ resembles the $T$ dependence
  of the isotropic scattering, while the anisotropy induces the
  variation from this result with similar $T$ dependence as observed in
  $R_H$ (see Fig. \ref{fig7}).  The magnetoresistance 
  strongly depends on the Fermi surface 
  shape (see inset and Fig. \ref{fig4}). All results are calculated
  for the fixed chemical potential. 
 }
\label{fig8a}
\end{figure}

Intra-layer magnetoresistance is like $R_H$ also sensitive to the
scattering anisotropy as is shown in Fig. \ref{fig8a} and in addition
shows $T$ dependence also for the isotropic scattering ($T_c=0$ case). This
can be traced back to its  proportionality to $(\omega_c \tau)^2$ 
dependence \cite{hussey96,sandeman01} for isotropic scattering, while
the proportionality factor strongly depends on the Fermi surface shape
(see the inset
in Fig. \ref{fig8a}).

\begin{figure}[htb] 
\centering 
\includegraphics[width = 0.3\textwidth, angle=-90]
  {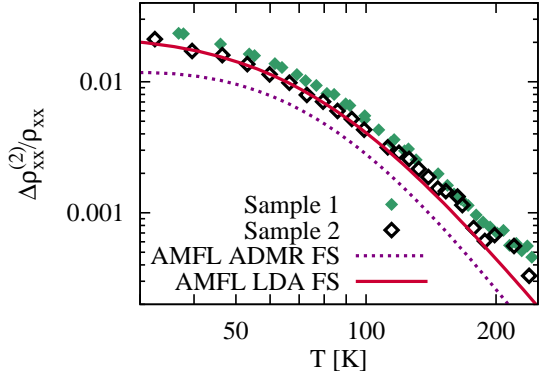}\\
  \caption{
  Comparison between the measured intra-layer magnetoresistance $\Delta
  \rho_{xx}^{(2)}/\rho_{xx}$ at $B_z=10$ T for two samples
  \cite{hussey96} and the result of the AMFL model for $T_c=25$
  K. AMFL results are calculated for the ADMR and LDA Fermi surfaces
  and agree qualitatively with the measured data.  
 }
\label{fig9a}
\end{figure}

Comparison of our calculations with experimental data for $T_c=25$ K
(Ref. \onlinecite{hussey96}) is shown in Fig. \ref{fig9a}. The
calculated magnetoresistance is in qualitative agreement with the
experimental data. Use of the LDA Fermi surface give quantitative
agreement.  However, considering the strong sensitivity of the
magnetoresistance to the small changes in the scattering anisotropy
\cite{zheleznyak99} or of the Fermi surface shape the comparison is
good.
  Previously it was pointed out that the cold spot model
\cite{ioffe98} cannot describe the intra-layer magnetoresistance of
underdoped and optimally doped cuprates. While our model is applicable
to the overdoped regime, it cannot describe the optimally doped or
underdoped regime, as already mentioned in \ref{sec_dcconductivity},
presumably due to the emergence of the pseudogap or other new physics
not included in our model.

\subsection{Modified Kohler's rule}

\begin{figure}[htb] 
\centering 
\includegraphics[width = 0.3\textwidth, angle=-90]
  {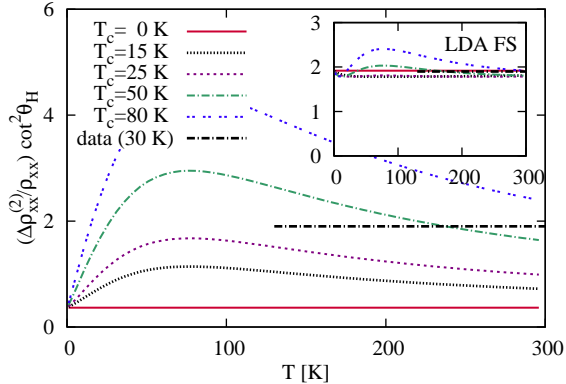}\\
  \caption{ Plot to test modified Kohler's rule. 
 Temperature dependence of 
    $(\Delta \rho_{xx}^{(2)}/\rho_{xx})\cot ^2\theta_H$, calculated
    for the 
    AMFL model, is shown for various $T_c$. Results show weak $T$ dependence,
    which is in agreement with the experimental observations shown
    with dash-dotted black line (ratio $\sim 1.9$)
    \cite{tyler98}. Results were calculated with the ADMR Fermi surface
    and depend strongly on the shape of the Fermi surface.  In the
    inset we show results for the LDA Fermi surface, which show weaker
    variation with $T$ and agree quantitatively with experimental
    data.  }
\label{fig10a}
\end{figure}

It has been observed that in underdoped and optimally doped cuprates
Kohler's rule 
\cite{mckenzie98a}, which states 
that the $\Delta\rho_{xx}^{(2)}/\rho_{xx}$ is a function of
$B/\rho_{xx}$, is strongly violated \cite{harris95} and therefore two
different scattering rates or anisotropic scattering needs to be
introduced. Furthermore, it has been realized that $(\Delta
\rho_{xx}^{(2)}/\rho_{xx})\cot ^2\theta_H$ is fairly constant with
temperature \cite{harris95} (modified Kohler's rule), which was argued
\cite{harris95} to support the separation of lifetimes picture put
forward by Anderson and co-workers, 
while the anisotropic scattering is inadequate and predicts too large
magnetoresistance \cite{ioffe98, sandeman01} (at least for optimal
doping). In Fig. \ref{fig10a} we show AMFL results for $(\Delta
\rho_{xx}^{(2)}/\rho_{xx})\cot ^2\theta_H$, 
 which show only weak $T$ dependence for $T>100$ K in the strongly overdoped regime
 in agreement with experiment. This supports the claims \cite{zheleznyak99,
  hussey03a} that anisotropic scattering can describe the weak $T$
dependence of this ratio.
However, the extent of the $T$ dependence seems to depend strongly on
the shape of the Fermi surface and is smaller for more square-like
Fermi surfaces. For example, we obtain quantitative agreement with
experimental data, if we use the LDA Fermi surface (see inset in
Fig. \ref{fig10a}).
 Support for the modified Kohler's rule can be
found also in the approximate $T^2$ dependence of $(\Delta
\rho_{xx}^{(2)}/\rho_{xx})^{-1/2}$ which is shown in Appendix
\ref{sec_mr_appendix}. 

\section{ Comparison with microscopic models}
\label{sec_Comparisonwithmicroscopicmodels}

It is a challenge for microscopic theory to quantitatively describe
the observed temperature and doping 
dependence of transport properties or equivalently the
proper $T$, $p$ and $\phi$
dependence of the self-energy in the overdoped cuprates. In this
section we compare our model self-energy to results from several
microscopic theories  in order to evaluate their
potential for a successful description of the various experimental data.  

A weak coupling treatment of the Hubbard model 
can produce an anisotropic
scattering rate of similar frequency and angular dependence to our model. 
The anisotropic MFL
component arises from a nesting of the Fermi surface in the anti-nodal
regions \cite{roldan06} or from proximity to a van Hove singularity
\cite{roldan06,kastrinakis05}. However, for the latter case the
resulting scattering rate would have the  opposite doping dependence and
would appear only at a higher 
temperature than that experimentally observed for Tl2201,
since the van Hove singularity would reach
the Fermi surface for dopings larger than in
the highly overdoped regime. 
Hence, the anisotropic MFL term can only arise from the nested parts
of the Fermi surface which produce a particle-hole susceptibility
similar to that found in one dimension.
Hence, the scattering is essentially arising from particle-hole
excitations with a high-frequency cutoff of the order of the band width.

A functional renormalization group treatment of the Hubbard model
\cite{ossadnik08} (for a review see 
Ref. \onlinecite{rice12}) shows a $T$, $p$ and $\phi$ dependence of the
scattering rate in qualitative agreement with ADMR
\cite{abdel06,abdel07} and our self-energy model. 
However, it predicts an order of magnitude
smaller anisotropic scattering rate than observed in experiment, while
it gives the correct order of magnitude 
for the isotropic scattering ($\propto
T^2$) in agreement with our self-energy model (see supplemental material
of Ref. \onlinecite{kokalj11}). 

The Hidden Fermi liquid (HFL) theory by Casey and Anderson
\cite{anderson06,casey11} uses a Gutzwiller projection of the Fermi
liquid wave function. 
However the scattering rate predicted by HFL has a linear $T$
dependence only for temperatures above $T \sim 400$ K, in strong contrast to the ADMR
measurements \cite{abdel06}, where the $T$ linear term is observed
even for $T< 60$ K \cite{kokalj11}.  Furthermore, within the HFL
theory the anisotropic scattering emerges solely as a consequence of
anisotropy of the Fermi momentum and of the Fermi velocity on the
Fermi surface
\cite{casey,casey11}. LDA calculations
\cite{peets07} show a weaker anisotropy and with the opposite doping
dependence than that needed in HFL to capture the experimentally observed
scattering rates \cite{kokalj11}.

Cluster dynamical mean field theory (CDMFT) can also calculate
scattering rates at different parts of the Fermi surface. Results
presented in Ref. \onlinecite{gull10} and obtained with a Hubbard model
with $t'/t=-0.15$ and $U=7t$ reveal qualitatively similar behaviour to ADMR
and to our model self-energy. For higher dopings CDMFT gives an isotropic
scattering rate, which becomes more anisotropic  (stronger scattering
in antinodal direction) and stronger with decreasing doping. However,
due to limitations of the quantum Monte Carlo method CDMFT is
currently limited to $T>0.05 t \sim 200$ K, which is above the most
interesting experimental regime. Quantitative comparison with our
self-energy model shows, that CDMFT \cite{gull10} predicts at $T=200$ K
a smaller isotropic part, by a factor $\sim 2.5$.
 Comparison of the
anisotropic part is complicated due to patch averaging in DMFT. 
However, the CDMFT self-energy \cite{gull10} has the
same order of magnitude as our model self-energy, at least at 
$T\sim 200$ K. Detailed quantitative comparison with the  CDMFT results is
given in Appendix \ref{sec_cdmft}.

Treatment of the $t$-$J$ model with the finite-temperature Lanczos
method (FTLM) \cite{jaklic00} yield results in good agreement with
several experimental data, including the optical conductivity and high $T$
resistivity. However, the temperature range of reliable results (due to
finite size effects) obtained with
FTLM is too high to address the low $T$ transport properties and in
particular the anisotropy in the scattering rate observed in ADMR.  

A large-$N$ expansion treatment of the $t$-$J$ model \cite{buzon10},
found a scattering rate with a similar temperature and angular
dependence as our model self-energy.  However, as optimal doping is
approached it also exhibits divergence of the anisotropic scattering
rate at low temperature, due to a $d$-density wave instability near
optimal doping.  This is qualitatively different from our model
self-energy.

Ioffe and Millis \cite{ioffe98} considered how superconducting
fluctuations could produce an anisotropic scattering rate. They
suggested that in the overdoped region the rate should scale with
$T^2$, but it should be kept in mind this depends on what assumptions
one makes about the temperature dependence and magnitude of the
superconducting correlation length. Superconducting fluctuations used
by Ioffe and Millis \cite{ioffe98} produce predominantly forward
scattering and so it is not clear to what extent they are effective in
transport.

Metzner and colleagues have been investigating d-density wave
fluctuations near a quantum critical point associated with a
Pomeranchuk instability.\cite{dellanna07} Their starting point was an
effective Hamiltonian which has a d-wave form factor built into
it. But this was motivated by earlier work \cite{Halboth00} on the
Hubbard model which found from renormalisation group flows
that strong forward scattering developed led to a Pomeranchuk
instability.  Although, this work reported an anisotropic scattering
rate that is linear in temperature it turns out that due to vertex
corrections the transport scattering time scales as $T^{4/3}$ and the
resistivity scales as $T^{5/3}$ (Ref. \onlinecite{dellanna09}).

In spite of all these theoretical studies the question remains whether
there is a simple explanation for the scattering in terms of a single
mechanism: e.g., antiferromagnetic, superconducting, or d-density wave
fluctuations. Furthermore, is there a smoking gun experiment which
could distinguish between these different contributions? For example,
they should have a different dependence on the magnitude of an
external magnetic field. We also note that a magnetic field couples
differently to spin and orbital degrees of freedom, and the former
contribution is dominant for fields parallel to the layers.

\section{Conclusions}
\label{sec_Conclusions}

In conclusion, we have shown that our model self-energy
can describe a wide range of experimental data on overdoped cuprates.
In earlier work we showed it could describe 
scattering rates deduced from ADMR,
the quasi-particle dispersion seen in ARPES,
 and effective masses deduced from specific heat and quantum magnetic
 oscillations  \cite{kokalj11}.  
Here, we have shown that neglecting vertex corrections the model can also
describe experimental data on electrical transport properties, including DC
conductivity, optical conductivity, Hall coefficient, and Hall angle. 

The small quantitative
discrepancies between the model and measured data at high frequencies
($>$1000 cm$^{-1}$) or higher $T$ ($>$300 K) could be 
reduced with 
application of a smoother high-frequency cutoff for the
self-energy, e.g. with the ``parallel
resistor'' formula \cite{hussey03a, hussey06}. 

The successful description of the experimental data
by our analysis shows that inclusion of vertex corrections is not
necessary at this level of approximation.
However, for the Hubbard
model on a square lattice it is claimed \cite{bergeron11,kontani08} that vertex
corrections are important in the optimal and underdoped
regimes.

Our results on the  DC resistivity show that in the
 overdoped regime the isotropic
scattering weakly depends on doping (or $T_c$), while the anisotropic
scattering increases super-linearly with increasing $T_c$ of decreasing
doping. Similar findings were obtained for LSCO in
Ref. \onlinecite{cooper09}. This highlights the fact that the doping
dependence of the DC resistivity in cuprates is generic and not so dependent
on material properties or Fermi-surface shape.

Such generic behaviour is not seen in the Hall effect, where 
for overdoped LSCO the Hall coefficient monotonically decreases with increasing
temperature and increasing doping \cite{narduzzo08}, with a sign
change for a doping $p \simeq 0.3$. 
This may be due to the proximity of the Fermi energy
 to the van Hove singularity in LSCO.

We have also shown that the main temperature
 dependence of the Hall coefficient $R_H$ comes from the temperature 
dependence of the self-energy anisotropy.
Our model was contrasted with the
Marginal Fermi liquid (MFL) model of Abrahams and Varma
\cite{abrahams03}, which 
consists of an anisotropic impurity scattering term and
an isotropic marginal
Fermi liquid term. This model was used to describe the $T$ dependence of
the Hall angle at optimal doping. However, their model cannot
describe the pronounced non-monotonic $T$ dependence of $R_H$  found
in overdoped Tl2201. It may be worth noting, that overdoped LSCO,
in contrast to Tl2201, shows a    monotonic $T$ dependence of $R_H$ and
so may be adequately described by the MFL model \cite{abrahams03}.

On the other hand, the observed $T^n$ dependence with $n\lesssim 2$ of
$\cot \theta_H$ is generic in the cuprates and has in combination with
the $T$-linear resistivity stimulated the proposal of more involved
theories. For example Anderson \cite{anderson91} suggested two types
of quasi-particles with different scattering rates. It was suggested
that, different scattering mechanism may be connected to the charge
conjugation properties of different currents\cite{coleman96}.
However, our analysis shows, that there is no need to evoke such
theories, since our anisotropic self-energy gives consistent
quantitative description of both $\rho_{xx}$ and $\cot \theta_H$. In
addition, we have shown that it also quantitatively describes the
frequency dependent conductivity, remarkably with no additional
fitting parameters and just using the parameters originally extracted
from ADMR \cite{kokalj11}.

Future work would could and should consider calculation of
thermoelectric transport properties such as the Seebeck coefficient
and Nernst signal using the same model self-energy.  In a
quasi-particle picture both of these transport coefficients contain
contributions from the energy dependence of the scattering time
\cite{behnia04,behnia09} and so may be sensitive to a marginal Fermi
liquid contribution to the self-energy.

The relevance of the model self-energy to electron doped cuprates
\cite{Armitage10} should also be investigated.  Recently it was
observed \cite{Jin11} that in the overdoped region of the phase
diagram the resistivity had a linear-in-temperature term which was
proportional to the superconducting $T_c$, as in the hole doped
cuprates considered here.

The broader significance of this work is that it shows that the
metallic state in the overdoped regime is not a simple Fermi liquid
and exhibits some physics which is similar to that found at optimal
doping [marginal Fermi liquid behaviour] and underdoping [anisotropic
Fermi surface properties with cold spots in the nodal directions].  A
significant challenge is to find a general phenomenological form of
the self-energy that with decreasing doping smoothly crosses over to a
form that describes the pseudogap state, such as the form proposed by
Yang, Zhang, and Rice \cite{yang06,rice12}.

\begin{acknowledgments}
This work was supported by an Australian Research Council Discovery Project
(DP1094395) and the EPSRC (UK).
We thank K. Haule, J. Merino, P. Prelov\v sek, B.J. Powell, and
J. Schmalian for helpful discussions. NEH would also like to
acknowledge a Royal Society Wolfson Research Merit Award. 
\end{acknowledgments}

\appendix

\section{Functions $f_H(\phi)$ and $f_{DC}(\phi)$}
\label{sec_fhphi}
Here we give explicit forms for the functions $f_H(\phi)$ and $f_{DC}(\phi)$
that appear
in Eq. (\ref{eq_rhdphi}). The function $f_H(\phi)$ can be readily obtained
from Eq. (\ref{eq_sigmaxy1_summary}) for $\sigma_{xy}^{(1)}$,
\eq{
f_H(\phi) =
\frac{e^3}{4\pi^2 d}
(-{\bf v}_{0,F}(\phi)\times \partial_\phi{\bf v}_{0,F}(\phi) )_z.
}
On the other hand, $f_{DC}(\phi)$ can be obtained from
Eq. (\ref{eq_dc_cond}) for $\RE \ \sigma_{xx}$, 
\eq{
f_{DC}(\phi) = 
\frac{ e^2}{4 \pi^2 d}  
\frac{k_F(\phi) v_{0,F}^2(\phi)}{v_{0,F,r}(\phi)}.
}

\section{Details on derivation of Hall conductivity}
\label{sec_detailsonderivation}

Here we give more details of the
 derivation of the expression for the Hall
conductivity, Eq. (\ref{eq_sigmaxy1_summary}), by starting with
Eq. (\ref{eq_sigmaxy1}), which is taken from Eqs. (2.7) and (3.36) in
Ref. \onlinecite{kohno88}. 
$J_x$ in  Eq. (\ref{eq_sigmaxy1}) represents the current vertex, which
in our approximation of neglecting 
vertex corrections equals $-e v_x$. The square brackets denote
\eq{
[A\tilde \partial_\mu B]=A\partial_{k_\mu} B - (\partial_{k_\mu} A) B,
}
which leads to 
\eq{
 [J_x \tilde \partial_y  J_y] = e^2 (v_x\partial_{k_y} v_y- v_y \partial_{k_y}
 v_x),
}
\eqa{
 [G^R\tilde \partial _x G^A] &=& G^R\partial_{k_x} G^A- \partial_{k_x}(G^R)
 G^A\nonumber\\
&=& G^R G^A (G^A (\hbar v_x +\partial_{k_x} \Sigma^A)\nonumber\\
&&-G^R(\hbar v_x
+\partial_{k_x} \Sigma^R)).
}
$G^{R(A)}$ represent the retarded (advanced) Green's function,
which may be written in terms of retarded (advanced) self-energy
$\Sigma^{R(A)}$.
With the use of $\Sigma^R=\Sigma^{A*}=\Sigma=\Sigma'+\im \Sigma''$ and 
$G^{R(A)}=1/(\omega-\epsilon_k-\Sigma^{R(A)})$ we can write
\eqa{
 &&[G^R\tilde \partial_x G^A]
 =\frac{1}{[(\omega-\epsilon_k-\Sigma')^2+(\Sigma'')^2]^2}\\
&&\times\big[
(\hbar v_x+\partial_{k_x} \Sigma')(-2 \im \Sigma'')+(-2\im \partial_{k_x}
\Sigma'')(\omega -\epsilon_k - \Sigma')
\big].\nonumber
}
The Hall conductivity can now be written as
\eqa{
&&\sigma_{xy}^{(1)} = \frac{-\im e^3 B_z}{2}
\sum_k
 \int \frac{d \omega}{2\pi}
(- \frac{\partial n_F(\omega)}{\partial \omega})\\
&&\times(v_x\partial_{k_y} v_y- v_y \partial_{k_y} v_x)
\frac{1}{[(\omega-\epsilon_k-\Sigma')^2+(\Sigma'')^2]^2}\nonumber\\
&&\times\big[
(v_x+\partial_{k_x} \Sigma')(-2 \im \Sigma'')+(-2\im \partial_{k_x}
\Sigma'')(\omega -\epsilon_k - \Sigma')
\big]. \nonumber
}
Since we neglected vertex corrections we should also neglect
$\partial_{k_x} \Sigma'$ in the above equation, which is the same as
neglecting the first correction to the vertex. For our
even-in-$\omega$ $\Sigma''$ first order vertex 
corrections ($v_x\to v_x+\partial_{k_x}\Sigma'$) turn out to be
negligible. Also the term with
$(\omega -\epsilon_k - \Sigma')$ may be 
neglected due to the strongly peaked and even-in-$\omega$ prefactor
$1/[(\omega-\epsilon_k-\Sigma')^2+(\Sigma'')^2]^2$. 
\eqa{
\sigma_{xy}^{(1)} &=& \frac{ e^3 B_z}{2}
\sum_k
 \int \frac{d \omega}{2\pi}
(- \frac{\partial n_F(\omega)}{\partial \omega}) \nonumber\\
&&\times\tilde v(k)
\frac{-2\Sigma''}{[(\omega-\epsilon_k-\Sigma')^2+(\Sigma'')^2]^2},
}
where 
\eq{
\tilde v(k) = v_x^2\partial_{k_y} v_y- v_x v_y \partial_{k_y} v_x.
\label{eq_tildevk}
}
The sum over $k$ may
be converted to an integral over  the first BZ, 
and the integral over $k_z$ can be performed due to the
quasi-two-dimensional nature of the system. The
 integral over $k_x$ and $k_y$ may be
decomposed into integrals over  $k_r$ 
[the radial direction from $(\pi,\pi)$, see Fig. \ref{fig4}]
and its azimuthal angle $\phi$. 
We are left with
\eqa{
\sigma_{xy}^{(1)} &=&
 \frac{e^3 B_z }{ (2\pi)^3 d}
\int d\phi
\int d \omega (- \frac{\partial n_F(\omega)}{\partial \omega}) \nonumber\\
&&\times\int dk_r k_r 
\hat v(k)
\frac{-2\Sigma''}
{[(\omega-\epsilon_k-\Sigma')^2+(\Sigma'')^2]^2}. 
}
In the next step we linearize the bare band dispersion close to the
Fermi surface
in the $k_r$ direction [see Eq. (\ref{eq_epsilon0k})]
and approximate 
\eqa{
&&\frac{1}
{[(\omega-\epsilon_k-\Sigma')^2+(\Sigma'')^2]^2}
\simeq\nonumber\\
&&\frac{\pi}
{2(-\Sigma''(\phi,\omega))^3 v_{0,F,r}(\phi) } 
\delta[k_r -\tilde k_r(\phi,\omega)],
}
with 
\eq{
\tilde k_r(\phi,\omega)=k_F(\phi)
+\frac{\omega -\Sigma'(\phi,\omega)}
{v_{0,F,r}(\phi)}. 
}
We further approximate $\tilde k_r(\phi,\omega) \sim
k_F(\phi)$, which we have checked numerically
results in an error of less than 2\% 
for the relevant band structures.
With this approximation the integral
over $k_r$ can be explicitly evaluated.
\eqa{
\sigma_{xy}^{(1)} =   
\frac{e^3 B_z }{2\pi^2 d}
\int d \phi
 \frac{k_F(\phi) \hat v (k_F(\phi),\phi) }
{v_{0,F,r}(\phi)}\nonumber\\
\times\int d\omega (- \frac{\partial n_F(\omega)}{\partial \omega}) 
\frac{1}{(-2\Sigma''(\phi,\omega))^2 }.
}
There is one further simplification regarding the "velocity" term that
can be done. Using the symmetry $\sigma_{xy}^{(1)}=-\sigma_{yx}^{(1)}$
we can write Eq. (\ref{eq_tildevk})
\eqa{
\tilde v&\to& \frac{1}{2}[
v_x^2\partial_{k_y} v_y +v_y^2\partial_{k_x} v_x-v_y
v_x( \partial_{k_y}(v_x)+ \partial_{k_x}(v_y))]\nonumber\\
&=&\frac{1}{2}({\bf v}\times {\bf
e}_z)\cdot (v_y{\bf \nabla}v_x -v_x{\bf \nabla}v_y). 
}
Expressing $({\bf v}\times {\bf e_z})= v {\bf t}$, where ${\bf t}$ is
unit vector parallel to the Fermi surface, and using
\eq{
{\bf \nabla}v_x \cdot {\bf t}= \frac{{\bf \nabla}v_x \cdot {\bf t}
  dk_\parallel}{dk_\parallel}= \frac{\partial_\phi
  v_x(\phi)}{k_f(\phi)/\cos\theta}, 
}
where $\theta$ is the angle between the Fermi surface direction and direction
${\bf e}_\phi$ (perpendicular to $k_r$). Analysing in the same way
the $y$ term brings us to
\eq{
\tilde v\to \frac{1}{2}
v_{0,F}(\phi)
[ - {\bf v}_{0,F}\times \partial_\phi {\bf v}_{0,F}(\phi)
]_z
\frac{\cos\theta}{k_F(\phi)}.
}
Finally, using $v_{0,F,r}(\phi)=\cos \theta v_{0,F}(\phi)$ cancels
$\cos \theta$ and we
can write our result as 
Eq. (\ref{eq_sigmaxy1_summary}).

\section{Effect of anisotropy on Hall effect}
\label{sec_effectofanisotropy}
\begin{figure}[htb] 
\centering 
\includegraphics[width = 0.3\textwidth, angle=-90]
  {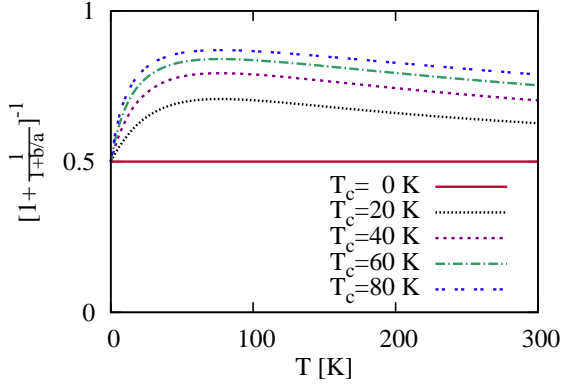}\\
  \caption{ Small temperature dependence of the
 factor $(1+\frac{1}{1+b/a})^{-1}$
    appearing in $\cot \theta_H$. This factor shows less than 10\%
    variation with $T$ from 50 to 300 K. Therefore the main $T$
    dependence of $\cot \theta_H$ comes from the isotropic scattering,
    which in our model self-energy is $\propto T^2$ (in
    agreement with experiment).  }
\label{fig9}
\end{figure}

Here we demonstrate how the anisotropy in the scattering rate
influences the Hall effect. In particular, we show with a simple
example that the $T$-dependence of the Hall coefficient $R_H$ is dominated
by $T$-dependent anisotropy, while, on the other hand, the
Hall angle
$\cot \theta_H$ and its $T$-dependence are dominated by the isotropic
scattering.
We start with the expressions for conductivities, which were used in
obtaining Eq. (\ref{eq_rhdphi}),
\eq{
\sigma_{xx}= \int d\phi f_{DC}(\phi) \frac{1}{-2\Sigma''(\phi,0)},
\label{eq_sigmaxx_intphi}
}
\eq{
\sigma_{xy}= \int d\phi f_H(\phi) \frac{1}{(-2\Sigma''(\phi,0))^2}.
\label{eq_sigmaxy_intphi}
}
In this simple approximation we neglect the $\phi$ dependence of
functions $f_H(\phi)$ and $f_{DC}(\phi)$ and exchange them with their
average values $\bar f_H$ and $\bar f_{DC}$. This is
feasible due to the much stronger anisotropy in the self-energy than in
 the $f$-functions.  Further on, we use a shorter notation for the two
 self-energy parts, $-\Sigma''_\textrm{FL}(0)=a$ and
 $-\Sigma_\textrm{AMFL}(0,0)=b$, which allows us to write
\eq{
-\Sigma''(\phi,0)=a +b \cos^2 (2\phi), 
}
where $a$ and $b$ are $T$ dependent. $a$ includes impurity scattering and
the FL
like part which is $\propto T^2$, while $b$ is due to the AMFL part and is
$\propto T$. With this approximation, integrals over $\phi$ in
Eq. (\ref{eq_sigmaxx_intphi}) and (\ref{eq_sigmaxy_intphi}) can be
explicitly performed and lead to 
\eq{
\sigma_{xx}= \pi \bar f_{DC} \frac{1}{a(1+\frac{b}{a})^{1/2}},
}
\eqa{
\sigma_{xy} =\frac{\pi}{4} \bar f_H
\frac{2+\frac{b}{a}}{a^2(1+\frac{b}{a})^{3/2}}. 
}
Expressing the Hall coefficient and Hall angle in this approximation
brings us to the final result of this section,
\eq{
R_H= \frac{1}{4\pi} \frac{\bar f_H}{\bar f_{DC}^2} 
\sqrt{1+\frac{b}{a}}
(1+\frac{1}{1+b/a}),
\label{eq_rh_anis}
}
\eq{
\cot \theta_H =4\frac{\bar f_{DC}}{\bar f_H} a
(1+\frac{1}{1+b/a})^{-1}.
\label{eq_cot_anis}
}
From Eq. (\ref{eq_rh_anis}) it is
evident that the Hall coefficient, and in particular its $T$
dependence, are dominated by
the $T$-dependent anisotropy $b/a$. On the other hand,
Eq. (\ref{eq_cot_anis}) reveals that the Hall angle $\cot \theta_H$ is
dominated by the isotropic scattering $a$, while the anisotropy effect
is strongly suppressed in the factor $(1+\frac{1}{1+b/a})^{-1}$.
The doping and $T$ dependence of $(1+\frac{1}{1+b/a})^{-1}$ for our model
self-energy are shown in Fig. \ref{fig9}. The effect of anisotropy can
be further increased or decreased by $\phi$ dependent $f_H$ or
$f_{DC}$, which can either increase or decrease the contribution from
 the AMFL part of the self-energy (or the antinodal part of the Fermi
surface). The effect of changing the $f$-functions by changing
the shape of the Fermi surface 
can for example be seen in Fig. \ref{fig5}.  Furthermore, the effect
of the anisotropy factor $(1+\frac{1}{1+b/a})^{-1}$ is to downturn the
$T^2$ dependence of $\cot \theta_H$ and make it more like $T^n$ with $n\leq
2$, which has in fact been observed (see inset in Fig. \ref{fig8} and
Refs. \onlinecite{ando99,fruchter07}.)

\section{Comparison of model self-energy with CDMFT}
\label{sec_cdmft}
\begin{figure}[htb] 
\centering 
\includegraphics[width = 0.3\textwidth, angle=-90]
  {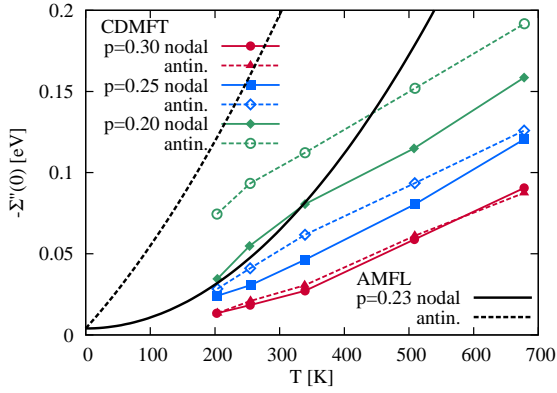}\\
  \caption{ Comparison of the imaginary part of the self-energy at
    zero frequency ($-\Sigma"(\omega=0)$) from CDMFT results
    (Ref. \onlinecite{gull10} Fig. 11) with our model
    self-energy. CDMFT results were obtained for different doping
    levels $p=0.3, 0.25,$ and $0.2$
    and for different patches on the Fermi surface (``nodal'' denotes
    nodal patch, while ``antin.'' denotes antinodal patch). CDMFT
    results are only available at higher temperatures ($T>200K$) due
    to limitations of the quantum Monte Carlo method used. Our model
    self-energy is most reliable at low $T$. (Experiments suggest it
    should become more linear for $T>200$ K (see Fig. \ref{fig1})).
    Most reliable comparison with CDMFT can therefore be done at
    $T=200$ K. At such $T$ CDMFT predicts a weaker isotropic
    self-energy (compare CDMFT $n=0.70$ and our nodal self-energy).
    Quantitative comparison for stronger anisotropies or lower doping
    is more difficult due to patch averaging in CDMFT. However, CDMFT
    predicts the correct trend with doping and order of the magnitude
    for the self-energy. Our antinodal self-energy was calculated with
    $T_c=60$ K, and the energy scale of CDMFT data was set with the
    hopping parameter $t_1=0.438$ eV.}
\label{fig10}
\end{figure}
\begin{figure}[!htb] 
\centering 
\includegraphics[width = 0.3\textwidth, angle=-90]
  {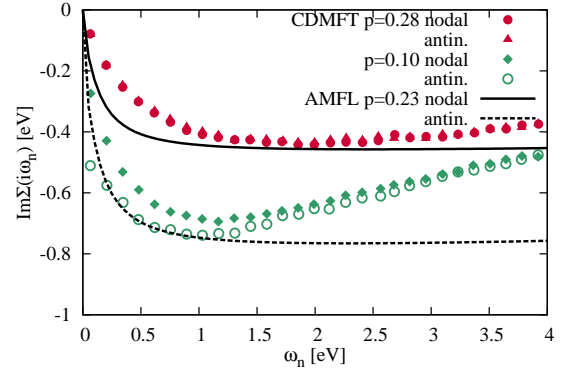}\\
  \caption{ Comparison of results for the Matsubara frequency
    dependence of $\IM\Sigma(i\omega_n)$ from CDMFT  
    (Ref. \onlinecite{gull10} Fig. 8) with our model self-energy. It is
    seen that the saturated value of the self-energy is similar in both
    results. The isotopic CDMFT result ($p=0.28$) predicts a smaller
slope at low frequencies (and thus are a quasi-particle weight closer
to one) than our model. 
Parameters for our model self-energy are the same as in
Fig. \ref{fig10} and the temperature corresponds to that of the CDMFT
calculation, $T=0.05 t_1 \sim 250 K$. }
\label{fig11}
\end{figure}

Here we show a quantitative comparison of our model self-energy with 
Cluster Dynamical Mean-Field Theory (CDMFT) calculations on the 
Hubbard model \cite{gull10}. Scattering at the Fermi surface or
$\Sigma''(\omega=0)$ is the most relevant quantity for explanation of
many transport data, which we analyze in this work. Our  model
$\Sigma''(\omega=0)$ at a doping level $p=0.3$ is compared with CDMFT results in
Fig. \ref{fig10}. 
Comparison of the dependence of the self-energy 
on the Matsubara frequencies on imaginary axis 
(see Fig. \ref{fig11}) 
can be done to avoid analytical continuation of CDMFT results. 
The slope at low frequencies
 ($\partial_{\omega_n} \IM\Sigma(i\omega_n)|_{\omega_n\to 0}$)
is related to the quasi-particle weight and mass renormalisation.

\section{Temperature dependence of intra-layer magnetoresistance}
\label{sec_mr_appendix}

\begin{figure}[!htb] 
\centering 
\includegraphics[width = 0.3\textwidth, angle=-90]
  {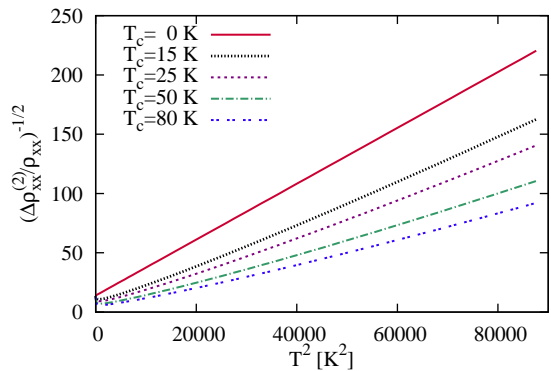}\\
  \caption{ Inverse square root of the intra-layer magnetoresistance,
    $(\Delta \rho_{xx}^{(2)}/\rho_xx)^{-1/2}$ vs. $T^2$. With this
    choice of the axes the $T^2$ behaviour of $(\Delta
    \rho_{xx}^{(2)}/\rho_xx)^{-1/2}$ becomes more apparent (at least
    at high $T$) and shows similar behaviour to $\cot
    \theta_H$. Therefore the ratio of the two is expected to show weak
  $T$ dependence and obeys the modified Kohler's rule as already
  discussed in the main text and shown in
  Fig. \ref{fig10a}. Curves are calculated with the ADMR Fermi surface and for
  several $T_c$s.}
\label{fig12a}
\end{figure}

Intra-layer magnetoresistance $\Delta \rho^{(2)}_{xx}/\rho_{xx}$
shows similarly to $\cot \theta_H$ (Fig. \ref{fig8}) $T^2$
temperature dependence, at least at higher $T$. This is shown in
Fig. \ref{fig12a} and implies the behaviour
according to the modified Kohler's rule.   

\FloatBarrier

\bibliographystyle{epj2}
\bibliography{ref_wr_pseudo.bib}

\end{document}